\documentclass[a4paper,num-refs]{oup-contemporary}

\journal{general}

\usepackage{graphicx}
\usepackage{siunitx}
\usepackage{caption}
\usepackage{subcaption}
\usepackage{multirow}
\usepackage{nameref}
\usepackage[section]{placeins}
\usepackage[many]{tcolorbox}    	
\usepackage{wasysym} 
\usepackage{array} 

\newcolumntype{L}[1]{>{\raggedright\arraybackslash}p{#1}}

\newcolumntype{R}[1]{>{\raggedleft\arraybackslash}p{#1}}

\title{Computational reproducibility of Jupyter notebooks from biomedical publications}

\author[1,2\authfn{1}\authfn{2}]{Sheeba Samuel}
\author[3,4,5\authfn{2}\authfn{3}]{Daniel Mietchen}
\affil[1]{Heinz-Nixdorf Chair for Distributed Information Systems,
Friedrich Schiller University Jena, Germany}
\affil[2]{Michael Stifel Center Jena, Germany}
\affil[3]{Ronin Institute, Montclair, New Jersey, United States}
\affil[4]{Institute for Globally Distributed Open Research and Education (IGDORE)}
\affil[5]{FIZ Karlsruhe~---~Leibniz Institute for Information Infrastructure, Berlin, Germany}

\authnote{\authfn{1}sheeba.samuel@uni-jena.de}
\authnote{\authfn{2}These authors contributed equally to this work}
\authnote{\authfn{3}daniel.mietchen@ronininstitute.org}

\newcommand{\initial}[1] {\lightning~#1}

\definecolor{main}{HTML}{5989cf}    
\definecolor{sub}{HTML}{cde4ff}     

\newtcolorbox{boxD}{
    colback = sub, 
    colframe = main, 
    boxrule = 0pt, 
    toprule = 3pt, 
    bottomrule = 3pt 
}

\papercat{Paper}

\runningauthor{Samuel et al.}

\begin{document}

\begin{frontmatter}
\maketitle
\begin{abstract}
\textbf{Background}
Jupyter notebooks facilitate the bundling of executable code with its documentation and output in one interactive environment, and they represent a popular mechanism to document and share computational workflows, including for research publications. 
The reproducibility of computational aspects of research is a key component of scientific reproducibility but has not yet been assessed at scale for Jupyter notebooks associated with biomedical publications.

\textbf{Approach}
We address computational reproducibility at two levels:
(1) Using fully automated workflows, we analyzed the computational reproducibility of
Jupyter notebooks 
associated with 
publications indexed in the biomedical literature repository PubMed Central. 
We identified such notebooks by mining the article's full text, trying to locate them on GitHub and attempting to re-run them in an environment as close to the original as possible. We documented reproduction success and exceptions and explored relationships between notebook reproducibility and variables related to the notebooks or publications.
(2) This study represents a reproducibility attempt in and of itself, using
essentially the same methodology twice on PubMed Central over the course of two years, during which the corpus of Jupyter notebooks from articles indexed in PubMed Central has grown in a highly dynamic fashion.

\textbf{Results}
Out 
of 27,271 
Jupyter notebooks from 2,660 
GitHub repositories associated with 3,467
publications,
22,578 
notebooks were written in Python, including 15,817 
that had their dependencies declared in standard requirement files and that we attempted to re-run automatically. 
For 10,388 
of these, all declared dependencies could be installed successfully, and we re-ran them to assess reproducibility. Of these, 1,203 
notebooks ran through without any errors, 
including 879 
that produced results identical to those reported in the original notebook, and 324 for which our results differed from the originally reported ones.
Running the other notebooks resulted in exceptions.

\textbf{Conclusions}
We zoom in on common problems and practices, highlight trends and discuss potential improvements to 
Jupyter-related workflows associated with biomedical publications.

\end{abstract}

\end{frontmatter}

\begin{keypoints*}
\begin{itemize}
\item We present a systematic attempt to automatically re-run Jupyter notebooks underlying research reported in articles indexed in PubMed Central.
\item The large majority of these notebooks could not be executed automatically, mostly due to issues with the documentation of dependencies.
\item The manuscript peer review process  often does not properly address the review of associated notebooks and would thus benefit from assistance by automated processes of the kind described here.
\end{itemize}
\end{keypoints*}



\section{Introduction} 
\label{sec:Intro}

Many factors contribute to the progress of scientific research, including the precision, scale, and speed at which research can be performed and shared and the degree to which research processes and their outcomes can be trusted \citep{siebert2015point,contera2021communication}. This trust, in turn, and the credibility that comes with it, are a social construct that depends on past experience or proxies to it \citep{gray2012understanding,kroeger2018scientific,jamieson2019Signaling}.
A good proxy here is reproducibility, at least in principle \citep{hsieh2018enhancing}: if a study addressing a particular research question can be re-analyzed independently and that analysis leads to the same conclusions as the original study, then these conclusions can generally be more trusted than if the conclusions differ between the original and the reproducibility study.

In the following sections, we provide a detailed description of our study. The \nameref{sec:Methods} section covers the techniques and the workflows employed to study the
reproducibility of Jupyter notebooks from GitHub repositories mentioned in PubMed Central publications. We then describe what we found in the \nameref{sec:Results} section. The \nameref{sec:Discussion} section 
contextualizes the results and
delves into the limitations and implications of our study. Finally, in \nameref{sec:Conclusions}, we summarize the key aspects of this article.

\subsection{Reproducibility issues in contemporary research}
\label{sec:Reproducibility-issues}

Over recent years, the practical reproducibility of published research has come into focus and turned into a research area in and of itself 
\citep{peng2015thereproducibility, samuel2021understanding}.
As a result, systematic issues with reproducibility have been the subject of many publications in various research fields as well as 
prominent mentions 
in the mass media \citep{theeconomist2013trouble}.
These research fields range from psychology \citep{simmons2011false} to cell culture \citep{hussain2013reproducible,bairoch2018cellosaurus} to ecology \citep{kelly2019rate}, geosciences \citep{ledermann2021towards}, open-source hardware \citep{Antoniou2021identifying} and beyond and include domains in which software plays a central role, such as health informatics \citep{coiera2018does}, 
human-computer interactions \citep{hinsen2018verifiability}, 
artificial intelligence \citep{hutson2018artificial,detlefsen2022torchmetrics}, software engineering \citep{shepperd2018role} and
research software \citep{crick2017reproducibility}. 
This is often framed in terms of a ``reproducibility crisis'' \citep{baker20161500}, though that may not necessarily be the most productive approach to addressing the underlying issues \citep{hunter2017the,fanelli2018opinion,guttinger2020limits}. In more practical terms,  
``appropriate workflow documentation is essential''~\citep{napflin2019genomics}, which includes
capturing appropriate metadata~\citep{leipzig2021role}.


\subsection{Terminology}
\label{sec:Terminology}
Within this broader context, distinctions between replicability, reproducibility, and repeatability are often important or even necessary \citep{meng2020reproducibility} but not consistently made in the literature \citep{plesser2017reproducibility}.
A potential solution to this confusion is the proposed distinction  \citep{goodman2016what} between 
\emph{Methods reproducibility} (providing enough detail about the original study that the procedures and data can be repeated exactly), 
\emph{Results reproducibility} (obtaining the same results when matching the original procedures and data as closely as possible) and
\emph{Inferential reproducibility} (leading to the same scientific conclusions as the original study, either by reanalysis or by independent replication).

In the following, we will concentrate on ``Methods reproducibility in computational research'', i.e.\ using the same code on the same data source. For this, we will use the shorthand ``Computational reproducibility''. In doing so, we are conscious that the ``same code'' can yield different results depending on the execution environment and that the ``same data source'' might actually mean different data if the data source is dynamic or if the code involves manipulating the data in a way that changes over time. 
We are also aware that the shorthand ``Computational reproducibility'' can also be 
used in other contexts,
e.g.
for
``Results reproducibility in computational research'' in cases where the algorithm described for the original study was re-implemented in a follow-up study. For instance, \citet{burlingame2021toward} were striving for \emph{Results reproducibility} when they re-implemented the PhenoGraph algorithm~-- which originally only ran on CPUs~-- such that it could be run on GPUs and thus at higher speed.
However, \emph{Results reproducibility} and \emph{Inferential reproducibility} are not the focus of our study~-- see~\cite{patel2022computing} for 
an example where these have been explored using Jupyter notebooks.




\subsection{Computational reproducibility in biomedical research}
\label{sec:Computational-Reproducibility}
In light of the reproducibility issues outlined above, there have been calls for better standardization of biomedical research software~-- see \citet{russell2018large} for an example.
In line with such standardization calls, a number of guidelines or principles
to achieve
methods reproducibility in several computational research contexts have been proposed. For instance, \cite{sandve2013ten}, \cite{gil2016toward} and \cite{Willcox2021ReSearchOps} laid out principles for reproducible computational research in general.
In a similar vein, \cite{gruning2018practical} and \cite{brito2020recommendations} looked at specifics of computational reproducibility in the life sciences, \cite{nust2020ten} explored the use of Docker~-- a containerization tool~-- in reproducibility contexts, 
and \cite{trisovic2022large} looked at the reproducibility of R scripts archived in an institutional repository, while
\cite{rule2019ten}, \cite{pimentel2019a} as well as \cite{wang2020restoring}, \cite{willis2020developing} and \cite{wang2020better} zoomed in on Jupyter notebooks, a popular file format for documenting and sharing 
computational workflows.
While most of these guiding documents are language agnostic, language-specific approaches to computational reproducibility have also been outlined, e.g.\  for Python
 \citep{halchenko2021datalad}.
 
 However, compliance with such standards and guidelines is not a given \citep{russell2018large,rule2018exploration,pimentel2021understanding}, so we set out to measure it specifically for Jupyter notebooks in the life sciences
 and to explore options to bridge the gap between recommended and actual practice.
 In order to do so, we mined a popular repository of biomedical fulltexts (PubMed Central) for mentions of
 Jupyter notebooks alongside mentions of a popular 
 repository for open-source software (GitHub). 
 
\subsection{PubMed Central} 
\label{sec:PMC}
PubMed Central (PMC)\footnote{\url{https://www.ncbi.nlm.nih.gov/pmc/}} is a literature repository containing full texts of biomedical articles.
At the time of writing, it contained about 9.2 million articles. 
Founded in the context of the Open Access mandate issued by the National Institutes of Health (NIH) in the United States \citep{roberts2001pubmed},
PMC is operated by the National Center for Biotechnology Information (NCBI), a branch of the National Library of Medicine (NLM), which is part of the NIH.
PMC hosts the articles using the Journal Article Tagging Suite (JATS), an XML standard, and makes them available for manual and programmatic access in various ways, of which we used the Entrez API \citep{sayers2010ageneral}.

\subsection{GitHub} 
\label{sec:GitHub}
GitHub\footnote{\url{https://github.com/}} is a website that combines git-based version control with support for collaboration and automation. It is a popular place for sharing software and developing it collaboratively, including for Jupyter notebooks \citep{rule2018exploration} and for code associated with research articles available through PubMed Central \citep{russell2018large}.

\subsection{Jupyter}
\label{sec:Jupyter}

Computational notebooks emerged in 1988 with the release of the proprietary software Mathematica \citep{wolfram88mathematica}, followed by Maple \citep{heck1993introduction} in 1989, which introduced a notebook-style graphical user interface.
In the past decade, the adoption of computational notebooks as
a computing environment in which code, code documentation and output of the code can be explored interactively
has greatly expanded, thanks to the rise of free and open-source platforms such as Project Jupyter \citep{kluyver2016jupyter,granger2021jupyter}\footnote{\url{https://jupyter.org/}}, RStudio \citep{team2015rstudio}\footnote{\url{https://posit.co/products/open-source/rstudio/}} and Pluto \citep{van_der_plas_2023_pluto}\footnote{\url{https://plutojl.org/}}.
Such notebooks facilitate data analysis, visualization, and collaboration, and they capture metadata about the steps performed, all of which contributes to the reproducibility and transparency of scientific research.

Jupyter notebooks
in particular
have become a popular mechanism to share computational workflows in a variety of fields \citep{kluyver2016jupyter}, including astronomy
\citep{randles2017using,wofford2019jupyter,patel2022computing} and biosciences \citep{schroder2019reproducible,malmstrom2019computational,Xue2021KinPred,Verwei2022Quantifying}. Here, we build on past studies of the reproducibility of Jupyter notebooks \citep{rule2018exploration,pimentel2019a,wang2021restoring} and 
automatically
analyze Jupyter notebooks available through GitHub repositories associated with publications whose full text is available through the biomedical literature repository PubMed Central.

\subsection{Jupyter and reproducibility}
\label{sec:Jupyter-reproducibility}
Jupyter notebooks can, in principle, be used to enhance reproducibility, and they are often presented as such, yet using them does not automatically confer reproducibility to the code they contain.
Several studies have been conducted in recent years to explore the reproducibility of Jupyter notebooks.
A recent one has investigated the reproducibility of Jupyter notebooks associated with five publications from the PubMed Central database \citep{schroder2019reproducible}. 
In their reproducibility analysis, they looked for the presence of notebooks, source code artifacts, documentation of the software requirements, and whether the notebooks can be re-executed with the same results. According to their results, the authors successfully reproduced only three of 22 notebooks from five publications.
Rule et al. \citep{rule2018exploration} explored 1 million notebooks available on GitHub. In their study, they explored repositories, language, packages, notebook length, 
and execution order, focusing on the structure and formatting of computational notebooks. As a result, they provided ten best practices to follow when writing and sharing computational analyses in Jupyter notebooks \citep{rule2019ten}.
Another study \citep{pimentel2021understanding} focused on the reproducibility of 1.4 million notebooks collected from GitHub. It provides an extensive analysis of the factors that impact reproducibility based on Jupyter notebooks.
Chattopadhyay et al. \citep{chattopadhyay2020s} reported on the results of a survey 
conducted among 156 data scientists on the difficulties when working with notebooks.
Other studies focus on best practices with respect to writing and sharing Jupyter notebooks \citep{rule2019ten, pimentel2021understanding, willis2020developing, wang2020better}.
As a result, tools have been developed to support provenance and reproducibility in Jupyter notebooks \citep{chirigati2013reprozip, boettiger2015Docker, samuel2018provbook, jupyter2018binder,kerzel2023mlprovlab}. 
Cases where Jupyter notebooks have played a key role in some actual reproducibility attempts have also begun to appear in the literature. For instance, 
a Jupyter notebooks 
were
assembled
in the context of assessing the reproducibility of the first images of black holes~\citep{patel2022computing} and
as part of a published correction in stem cell research~\citep{baker2019quantification},
whereas 
an epidemiological
paper was published with a Jupyter notebook that enabled  others to reproduce the computational workflows, ultimately leading to the retraction of the original work, as detailed in \cite{meyerowitz2021impact}.

\subsection{Environmental footprint} 
\label{sec:Environmental-footprint}
Computations ultimately require physical resources, and 
awareness is growing that
both the production and the use of these resources can have a considerable environmental footprint \citep{lannelongue2021green,loic2023greener}. 
The more reproducible some workflows become, the more accurately their environmental footprint can be assessed \citep{taddeo2021artificial}. 
This can then lead to an optimization of the environmental footprint, especially since it often correlates with the financial footprint of using computational resources \citep{schwartz2020green}.
One of our aims in this study is thus to get an overview of the contribution of Jupyter-based workflows to the environmental footprint of biomedical research involving computation. This is in line with 
the need for humanity to act within earth system boundaries \citep{rockstrom2023safe}
and
the recommendation in \citet{lannelongue2021ten} to integrate routine environmental footprint assessment into research practice.
For practical reasons, we focus here on the carbon dioxide production, ignoring other 
greenhouse gases~\citep{montzka2011non}
as well as other
components of the ecological footprint~-- such as the use of water~\citep{li2023making}~-- or trade-offs between algorithmic performance and environmental impact, which have just begun to be explored in a systematic fashion~\citep{kaddour2023train}.

\section{Methods}
\label{sec:Methods}


The methodology employed here is largely identical to that reported in
our 2022 preprint \citep{samuel2022computational}, with the main difference being that here, we report on a re-run of our pipeline, rather than the initial run that was the focus there.

When reporting on our methodology and results,
we will 
thus provide
the values from the 2023 re-run and~-- whenever 
feasible~-- 
complement that (in parentheses and prepended with a lightning symbol, \lightning ) with the values from the original 2021 run, to help assess trends in this highly dynamic space. Likewise, all figures presented here are based on data from the re-run. The 2021 values, tables and figures are available via the preprint \citep{samuel2022computational}.


\subsection{Pipeline} 
\label{sec:Pipeline}

\begin{figure*}[ht!]
\centering
\includegraphics[width=1.0\textwidth]{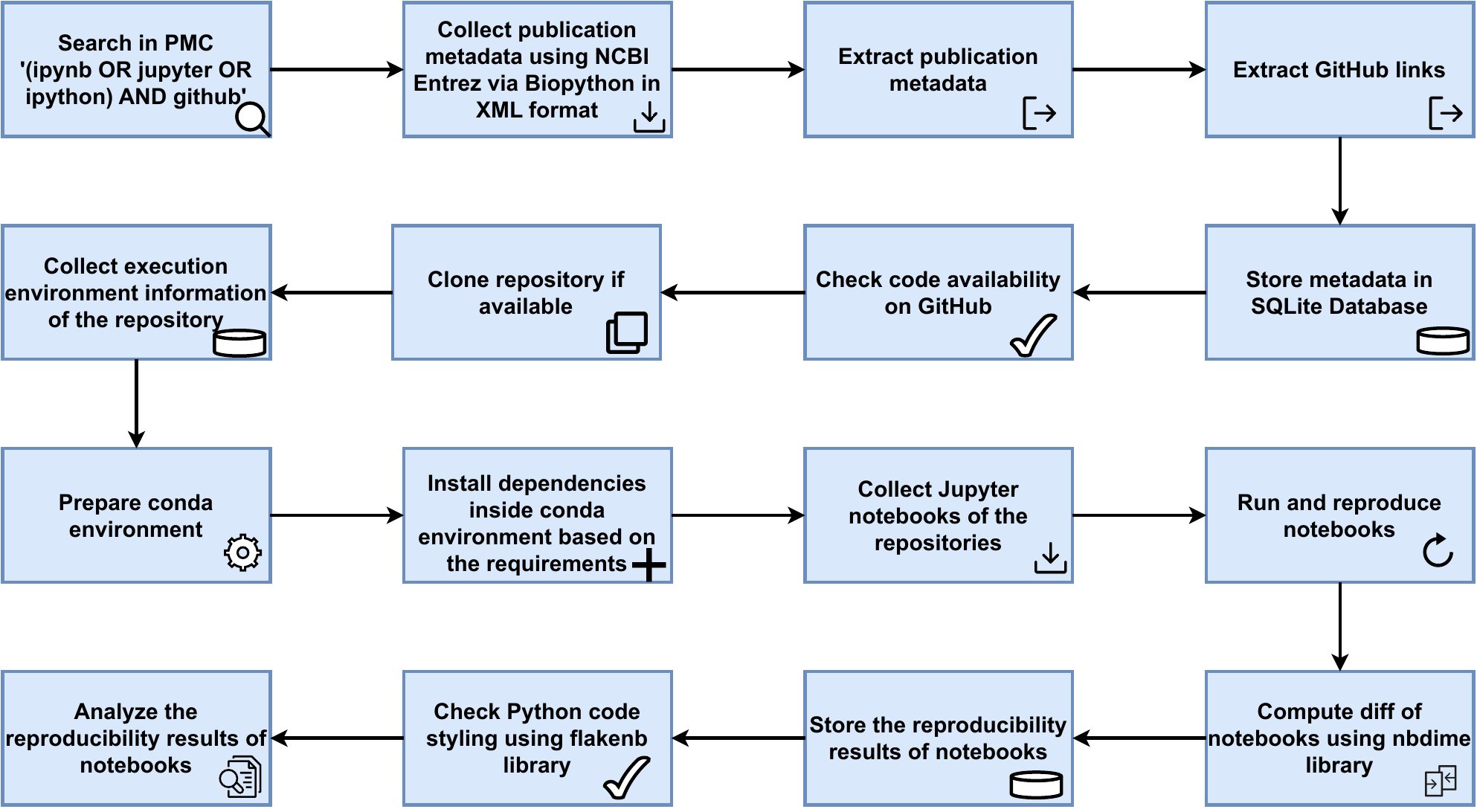}
\caption[Workflow]{
Fully automated workflow used for assessing the reproducibility of Jupyter notebooks from publications indexed in PubMed Central: the PMC search query resulted in a list of article identifiers that were then used to retrieve the full-text XML, from which publication metadata and GitHub links were extracted and entered into an SQLite database. If the links pointed to valid GitHub (RRID:SCR\_002630) 
repositories containing valid Jupyter notebooks, then metadata about these were gathered, and 
Python-based notebooks were run with all identifiable dependencies, and their results analyzed with respect to the originally reported ones.
}
\label{fig:Figure_MethodsWorkflow}
\end{figure*}

In this section, we describe the key steps of the pipeline we used for assessing the reproducibility of Jupyter notebooks (RRID:SCR\_018315) 
available via GitHub (RRID:SCR\_002630) repositories
extracted from the full text of
publications 
indexed in
PubMed Central (RRID:SCR\_004166). 
The driver file 
for running the workflow was {\it r0\_main.py}\footnote{For the location of these files, see Section \nameref{sec:Dataavailability}.}, and the 
driver notebook for the analysis of the collected data was {\it Index.ipynb}\footnote{We chose this file name before our 
analysis made us aware that it is a common name for Jupyter notebooks~-- see Section 
\nameref{sec:NotebookNaming}
for details.}.
Figure \ref{fig:Figure_MethodsWorkflow} 
provides a conceptual overview of the workflow used in this study.

\subsubsection{PMC search} 
\label{sec:pmcsearch}
We used the \textit{esearch} function to search PMC for Jupyter notebooks on 27th March 2023 (\initial{24\textsuperscript{th} February, 2021}).
We looked for publications that mentioned GitHub together with either the string ``Jupyter'' or some closely associated ones, namely ``ipynb'' (the file ending/extension of Jupyter notebooks) or ``IPython'' (the name of a precursor to Jupyter).
The search query used was ``(ipynb OR jupyter OR ipython) AND github''.
Based on the primary PMC IDs received from the \textit{esearch} utility, we retrieved records in the XML format using the \textit{efetch} function and collected the publication metadata from PMC \citep{roberts2001pubmed} using NCBI Entrez utilities via Biopython \citep{cock2009biopython}.

\subsubsection{Metadata extraction}
\label{sec:extraction}
In the next step, we processed the XML fetched from PMC.
We used an SQLite database\footnote{\url{https://www.sqlite.org}} for storing all the data related to our pipeline.
We collected information on journals and articles.
We first extracted information about the journal.
For this, we created a database table 
for the journal and extracted the ISSN\footnote{\url{https://www.issn.org/}} (International Identifier for serials), the journal title, the NLM's (National Library of Medicine) abbreviated journal title, and the ISO\footnote{\url{https://www.iso.org}} (International Organization for Standardization) abbreviation.
\newpage

We then created a database table for the articles and populated it with article metadata.
The metadata includes the article name, Pubmed ID, PMC ID, Publisher id and name, DOI, subject, the dates when the article was received, accepted, and published, the license, the copyright statement, keywords, and the GitHub repositories mentioned in the publication.
For each article, we also extracted the associated Medical Subject Headings (MeSH terms)\footnote{\url{https://www.ncbi.nlm.nih.gov/mesh}}, of which they typically have several.
These terms are assigned to articles upon indexing in the PubMed database. PubMed is a database of abstracts, and it usually has an entry for articles indexed in its full-text companion, PubMed Central.
These MeSH terms are hierarchical, and we obtained the top-level MeSH term by querying the MeSH RDF API through SPARQL queries to the SPARQL endpoint\footnote{\url{https://id.nlm.nih.gov/mesh/sparql}}.  We then
aggregated them by top-level terms (amounting to 108 in our dataset) that served as a proxy for the subject areas of the article.

To extract the GitHub repositories mentioned in each article, we looked for mentions of GitHub links anywhere in the article, 
including the abstract, the article body, data availability statement, and supplementary information.
GitHub links were available in different formats.
We normalized them to the standard format `https://github.com/\{username\}/\{repositoryname\}'.
For example, we extracted the GitHub repository from nbviewer\footnote{\url{https://nbviewer.org/}} links and transformed its
representation to the standard format.
We excluded 682 (\initial{172}) GitHub links that mentioned only the username or organization name or GitHub Pages and not a specific repository name.
After preprocessing and extracting GitHub links from each article, we added the GitHub repositories to the database table for the corresponding articles.
Likewise, we linked the article's entry in the table to the journal where it was published.
We also collected information on the article authors in a separate 
author database table, extracted the first and last name, ORCID, email, and connected these data to the corresponding entries in the article table. 

Based on the GitHub repository name collected from the article, we checked whether these repositories were available at the original link or not. 
If the repository existed, we cloned it (ignoring branches, i.e.\ just taking the default one, which is usually called ``main'' for new repositories, or ``master'' for older ones) and collected information about the repositories using the GitHub REST API\footnote{\url{https://docs.github.com/en/rest/guides/getting-started-with-the-rest-api}}.
On that basis, we created a repository database table.
For each GitHub repository, an entry is created in the table and connected to the article where it is mentioned.
Additional information for each repository is also collected from the GitHub API. 
This includes the dates of the creation, updates, or pushes to the repository, and the programming languages used in each repository. 
Further information includes the number of subscribers, forks, issues, downloads, license name and type, total releases, and total commits after the respective dates for when the article was published, accepted, and received.
For each notebook provided in the repositories, we collected information on the name, nbformat, kernel, language, number of different types of cells, and the maximum execution count number.
We extracted the source and output of each cell for  further analysis.
Using Python Abstract Syntax Tree (AST)\footnote{\url{https://docs.python.org/3/library/ast.html}}, the pipeline extracted information on the use of modules, functions, classes, and imports.

\subsubsection{Notebook styling}
\label{sec:notebookstylingmethod}
After collecting the notebooks, we additionally ran a Python code styling check using the \textit{flakenb}\footnote{\url{https://github.com/s-weigand/flake8-nb}} library on the notebooks, since code styling consistency is a potential indicator for the extent of care that went into a given piece of software.
The \textit{flakenb} library is a tool for code style guide enforcement for notebooks.
It helps to check code against some of the style conventions in PEP 8\footnote{\url{https://www.python.org/dev/peps/pep-0008/}}, a style guide for Python code.
\textit{flakenb} 
provides an \textit{ignore} flag to ignore some specified errors.
In this study, we did not use this flag and collected all errors detected by the library.
For the styling of notebooks, we collected information on the pycode styling error code and description\footnote{\url{https://pycodestyle.pycqa.org/en/latest/intro.html}}.

\subsubsection{Computational environment setup}
\label{sec:envsetup}
We collected the execution environment information by looking into the dependency information declared in the repositories in terms of files like \textit{requirements.txt}, \textit{setup.py} and \textit{pipfile}.
After collecting all the required information for the execution of Python notebooks from the repositories, we prepared a conda\footnote{\url{https://docs.conda.io/en/latest/}} 
environment based on the python version declared in the notebook.
Conda is an open source package and environment management system which helps users to easily find and install packages and create, save, load and switch between environments.
The pipeline then installed all the dependencies collected from the corresponding files like \textit{requirements.txt}, \textit{setup.py} and \textit{pipfile} inside the conda 
environment. 
For the repositories that did not provide any dependencies using the above mentioned files, the pipeline executed the notebooks by installing all 
anaconda 
dependencies\footnote{\url{https://docs.anaconda.com/anaconda/packages/pkg-docs/}}.
Anaconda is a Python and R distribution which provides data science packages including \textit{scikit-learn}, \textit{numpy}, \textit{matplotlib}, and \textit{pandas}.

\begin{figure*}[ht!]
\centering
\includegraphics[width=1.0\textwidth]{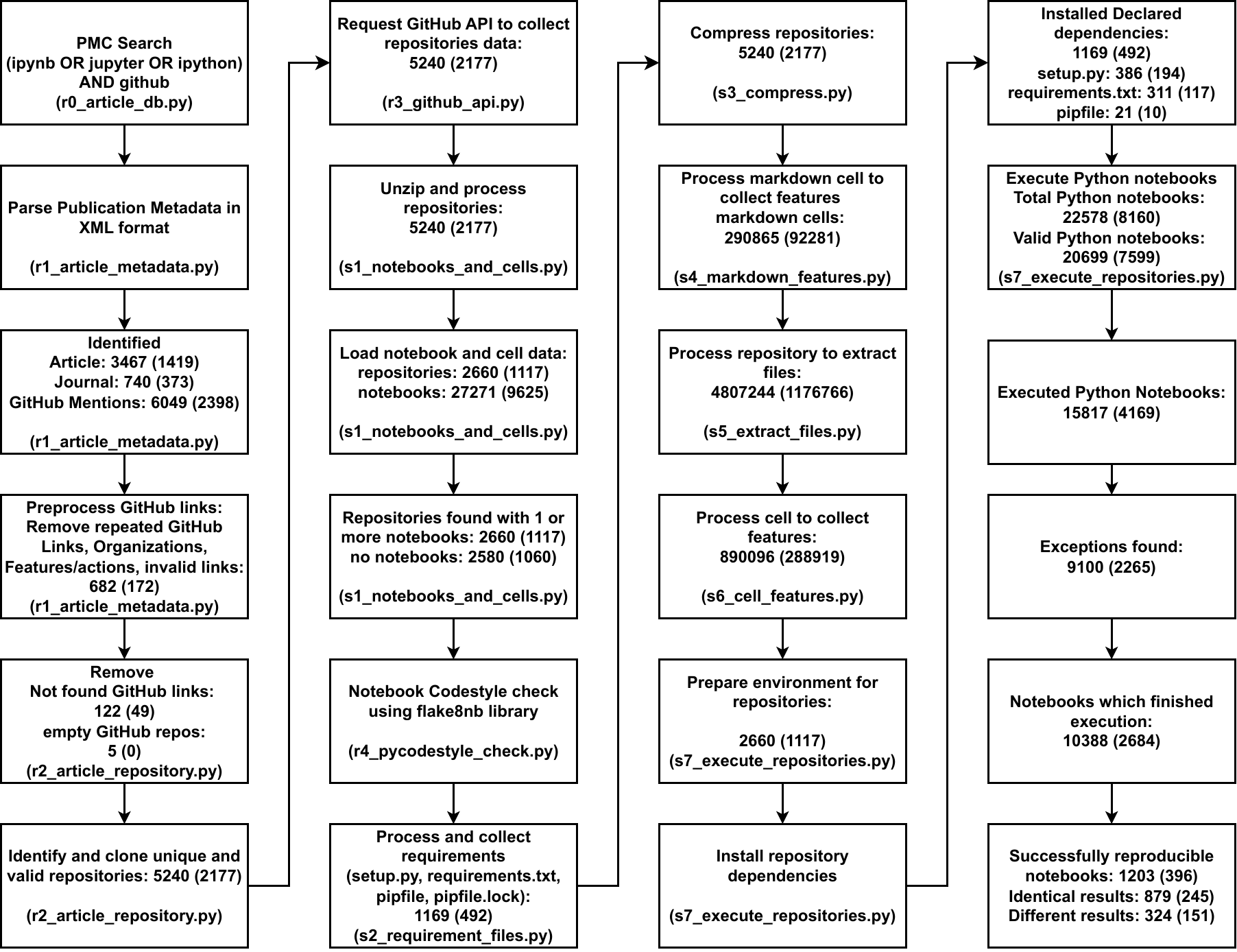}
\caption[Workflow]{
Key steps of the computational workflow used for the study, illustrated in a way that is partly inspired by the PRISMA flow diagram~\citep{page2021PRISMA}.
Each box contains a brief description of the corresponding step and the numbers of entities tracked at that step. The numbers given in parentheses indicate the results of the initial run of the pipeline in 2021 \citep{samuel2022computational}.
The name of the file containing the code for the respective step is indicated at the bottom of its box. 
}
\label{fig:Figure_PRISMA}
\end{figure*}

\subsubsection{Reproducibility pipeline and analysis tools}
\label{sec:pipelinetools}
After collecting and creating these data tables, we ran a pipeline to run the Jupyter notebooks contained in the GitHub repositories. 
The code for the pipeline is adapted from \citep{pimentel2019a, samuel2021reproducemegit}.
Hence, the method to reproduce the notebooks in this study is similar to \citep{pimentel2019a}.
The ReproduceMeGit \citep{samuel2021reproducemegit}~-- extended from \citep{pimentel2019a}~-- is a visualization tool for analyzing the reproducibility of Jupyter notebooks, along with provenance information of the execution.
ReproduceMeGit provides the difference between the results of the executions of notebooks using the nbdime\footnote{\url{https://github.com/jupyter/nbdime}} library.
These two tools provide the basis for our code for the reproducibility study.

\subsubsection{Code structure and visualization}
\label{sec:}
In this study, we use Jupyter notebooks for data computation and analysis. We created two sets of notebooks: one set (naming pattern N[0-9]*.ipynb) is focused on examining data pertaining to repositories and notebooks, while the other set (PMC[0-9]*.ipynb) is for analyzing data associated with publications in PubMed Central, i.e.\ for plots involving data about articles, journals, publication dates or research fields. 
The code used to generate each figure presented in this paper is available in these two sets of notebooks.
To facilitate data processing, essential Python libraries like \textit{numpy}\footnote{\url{https://numpy.org/}} and \textit{pandas}\footnote{\url{https://pandas.pydata.org/}} are employed. Additionally, for interactive data visualization, plotly\footnote{\url{https://plotly.com/python/}} (an open source graphing library for Python) is utilized alongside matplotlib.pyplot\footnote{\url{https://matplotlib.org/stable/tutorials/introductory/pyplot.html}}.

\subsection{Computation}
\label{sec:Computation}

\subsubsection{Initial run (2021)}
\label{sec:re-run-initial}

The pipeline outlined in Figure ~\ref{fig:Figure_PRISMA} was set up through the Friedrich Schiller University Ara Cluster\footnote{\url{https://wiki.uni-jena.de/pages/viewpage.action?pageId=22453005}} on a Skylake Standard Node (2x Intel Xeon Gold 6140 18 Core 2,3 GHz, 192 GB RAM).
This node has two CPUs, each with 18 cores, and 192 GB RAM in total.
The complete pipeline ran from 24\textsuperscript{th}-28\textsuperscript{th} February 2021 for a total of 117 hours and 52 minutes. 

\subsubsection{Re-run (2023)}
\label{sec:self-replication-run}
We then re-ran the entire pipeline using the same setup, except that 
we now allocated 128GB of memory to the task, when this was not specified in the intial run.
This took from 27 March till 9 Mai 2023, for a total of 43 days. 
Additional commits (1-16) are associated with various types of errors and updates, including external dependencies, deprecation issues, compatibility errors, conflicts in dependencies, empty or incomplete repositories, typo or missing checks, and additional support\footnote{\url{https://github.com/fusion-jena/computational-reproducibility-pmc/commits/main}}.
In the re-run of our study, we encountered various interruptions, including a power failure in the ARA cluster hosted by the University. We also had to deal with file storage system problems on April 25, 2023, as well as outages on April 27, 2023, and similar issues in January\footnote{\url{https://wiki.uni-jena.de/display/WIL/2023/04}}.
Some notebooks in certain repositories ran for several days, continuously producing `RequestException' errors in the logs without stopping, even though we had set a default timeout. As a consequence, we had to exclude them from our analysis. These interruptions, in general, increased the total run time of our pipeline.

\subsection{Environmental footprint estimation} 
\label{sec:Environmental-footprint-estimation}
We used 
the website \url{https://green-algorithms.org} v2.2 \cite{lannelongue2021green} 
that takes the hardware configuration, the total runtime and the location as input and then provides 
an estimate of the environmental footprint of the computation.
Our calculation does not include software development on our side or for any of our dependencies, nor test runs, figure generation or any other activity related to the project.


\section{Results} 
\label{sec:Results}

In this section, we present the results of our study 
analyzing the computational reproducibility of Jupyter notebooks from biomedical publications.

\subsection{General statistics of our study}
\label{sec:GeneralStatistics}

We extracted metadata from 3,467 (\initial{1,419}) publications from PubMed Central. 
These articles had been published in 740 (\initial{373}) journals and had 6049 (\initial{2398}) mentions of GitHub repository links.
At the time of data collection, 122 (\initial{49}\footnote{These 49 repositories were still inaccessible at the time of the re-run.}) GitHub repositories mentioned in the articles were not accessible, returning a ``page not found'' error instead. 
Out of 5,240 (\initial{2,177}) unique and valid GitHub repositories cloned, only 2,660 (\initial{1,117}) had 
at least
one 
Jupyter notebook.
From these repositories, a total of 27,271 (\initial{9625}) Jupyter notebooks were downloaded for further reproducibility analysis.
This dataset can be explored at various levels, e.g. articles, journals, GitHub repositories, Jupyter notebooks and any of their respective metadata dimensions, some of which we will highlight in the following.

\subsection{Research fields}
\label{sec:Research_fields}

\begin{figure}[!htb]
\includegraphics[width=1.05\linewidth]{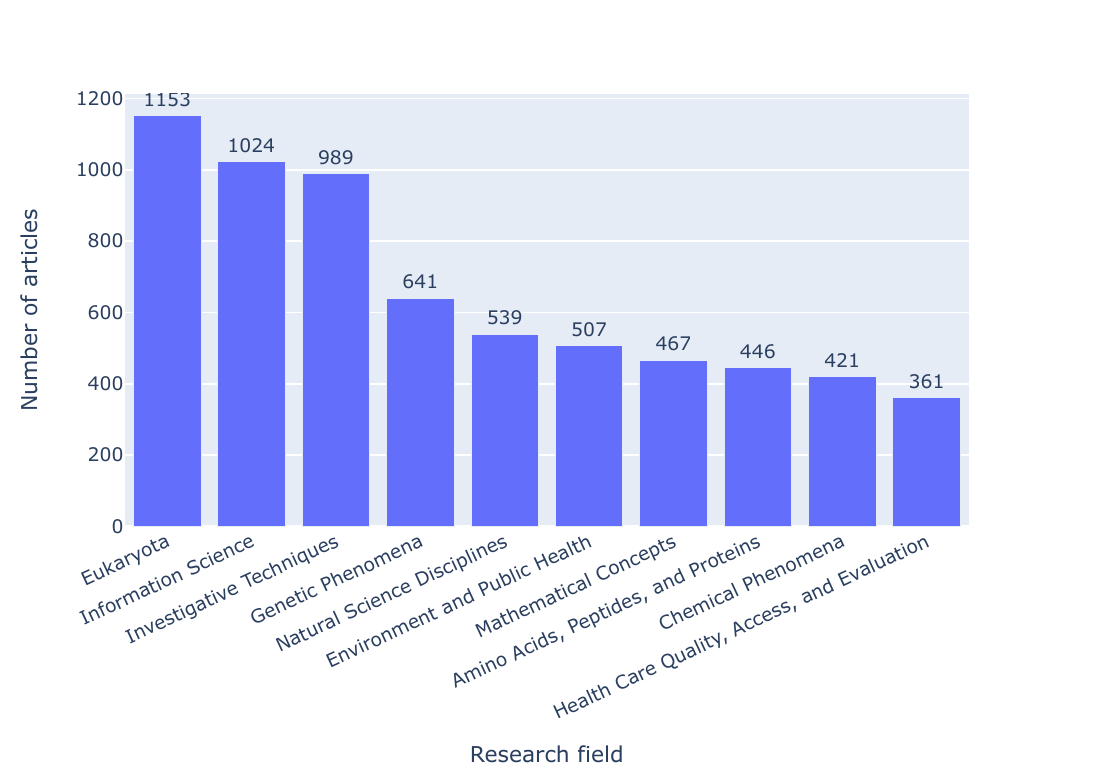}
\caption{
Full-text articles from PMC that mention GitHub repositories, grouped by 
top-level MeSH terms as a proxy for their research field. 
}
\label{fig:Figure_top_researchfields_with_articles}
\end{figure}
\begin{figure}[!htb]
\includegraphics[width=1.05\linewidth]{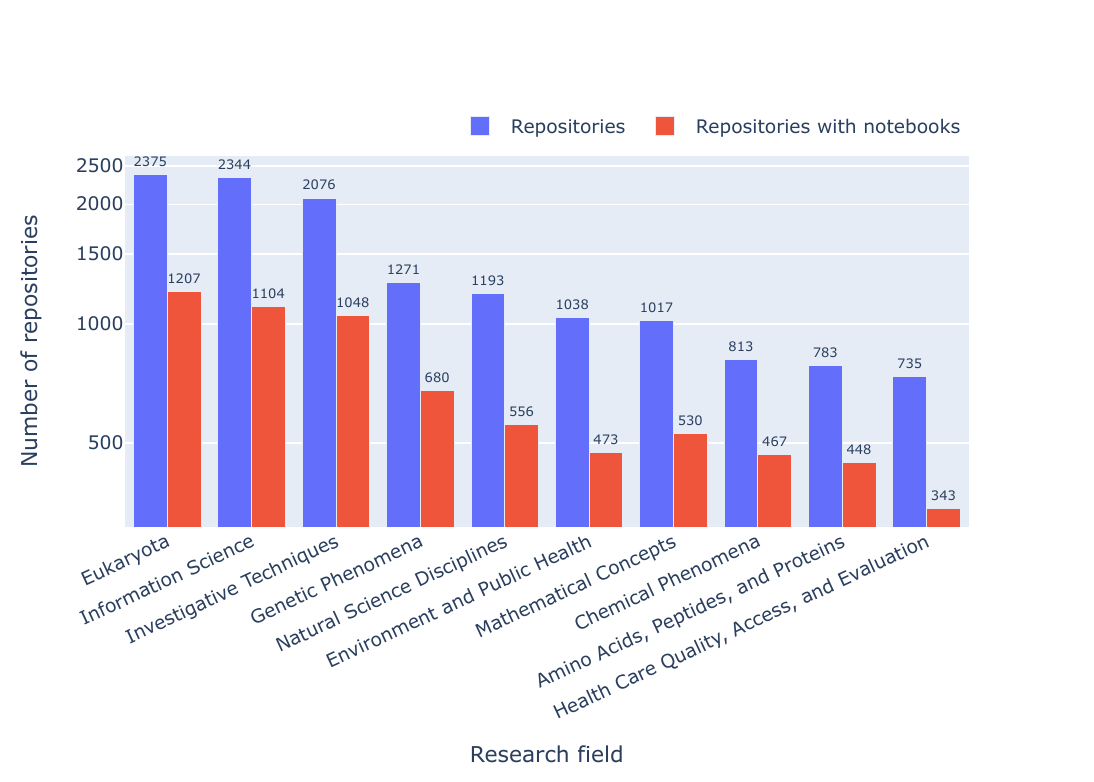}
\caption{
MeSH terms by the number of GitHub repositories mentioned in our corpus, highlighting (in red) those that contain at least one Jupyter notebook.}
\label{fig:Figure_top_researchfields_with_without_repositories_notebooks}
\end{figure}

Using MeSH terms as a proxy for research field, we can, for instance, rank fields by the number of PMC-indexed articles that mention GitHub repositories (cf.\ Figure \ref{fig:Figure_top_researchfields_with_articles}),
or aggregate the MeSH terms across articles and 
then filter by presence of Jupyter notebooks, thus ranking fields 
by number of 
mentioned GitHub repositories with or without Jupyter notebooks, as shown in Figure \ref{fig:Figure_top_researchfields_with_without_repositories_notebooks}.
MeSH terms can refer, for instance, to the object of study (e.g. Eukaryota) to the knowledge domain (e.g. Information Science) or notions of methodology (e.g. Investigative Techniques), and Figures 
\ref{fig:Figure_top_researchfields_with_articles}
and
\ref{fig:Figure_top_researchfields_with_without_repositories_notebooks}
highlight that our corpus contains a broad mix of these. Figure \ref{fig:Figure_top_researchfields_with_without_repositories_notebooks}
also illustrates that only about half of the mentioned GitHub repositories 
actually contained Jupyter notebooks, rather irrespective of the respective MeSH terms.

\begin{figure}[!htb]
\includegraphics[width=\hsize]{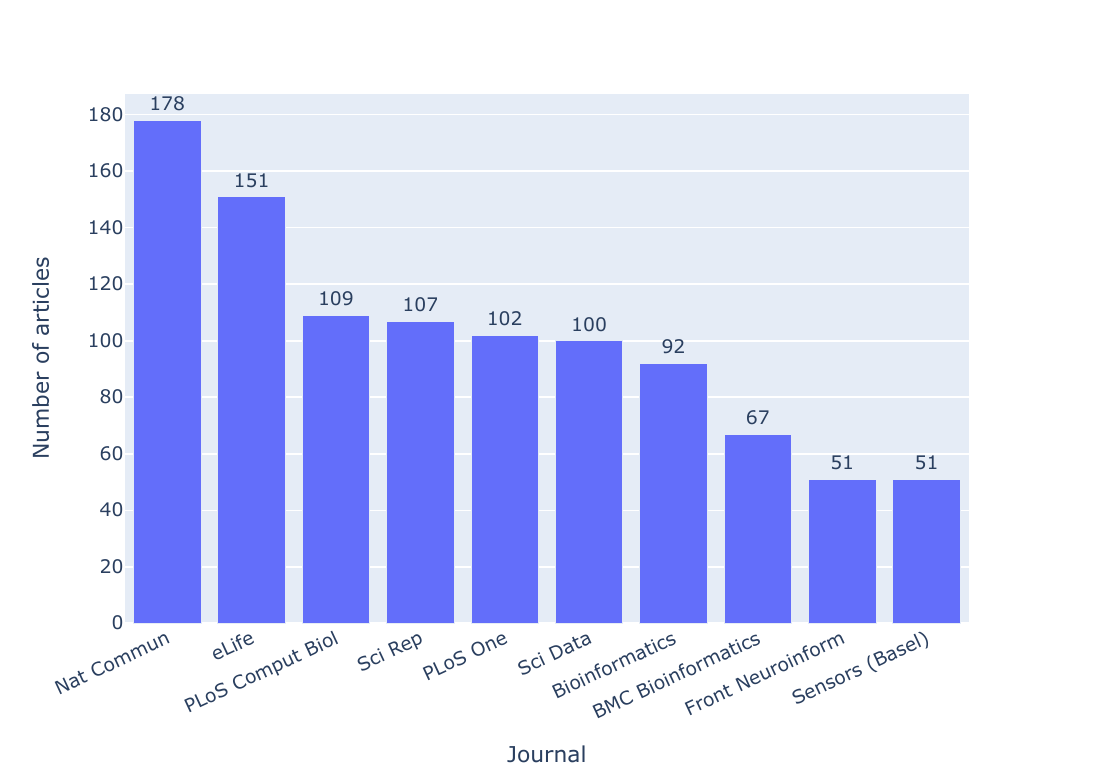}
\caption[Journals with the highest number of articles that had a valid GitHub repository and at least one Jupyter notebook.]{
Journals with the highest number of articles that had a valid GitHub repository and at least one Jupyter notebook. In the figures, journal names are styled as in the XML files we parsed, e.g. (``PLoS Comput Biol''). In the text, we use the full name in its current styling, e.g.  ``PLOS Computational Biology''. 
}
\label{fig:Figure_top_journals_with_articles}
\end{figure}

\begin{figure}[!htb]
\includegraphics[width=\hsize]{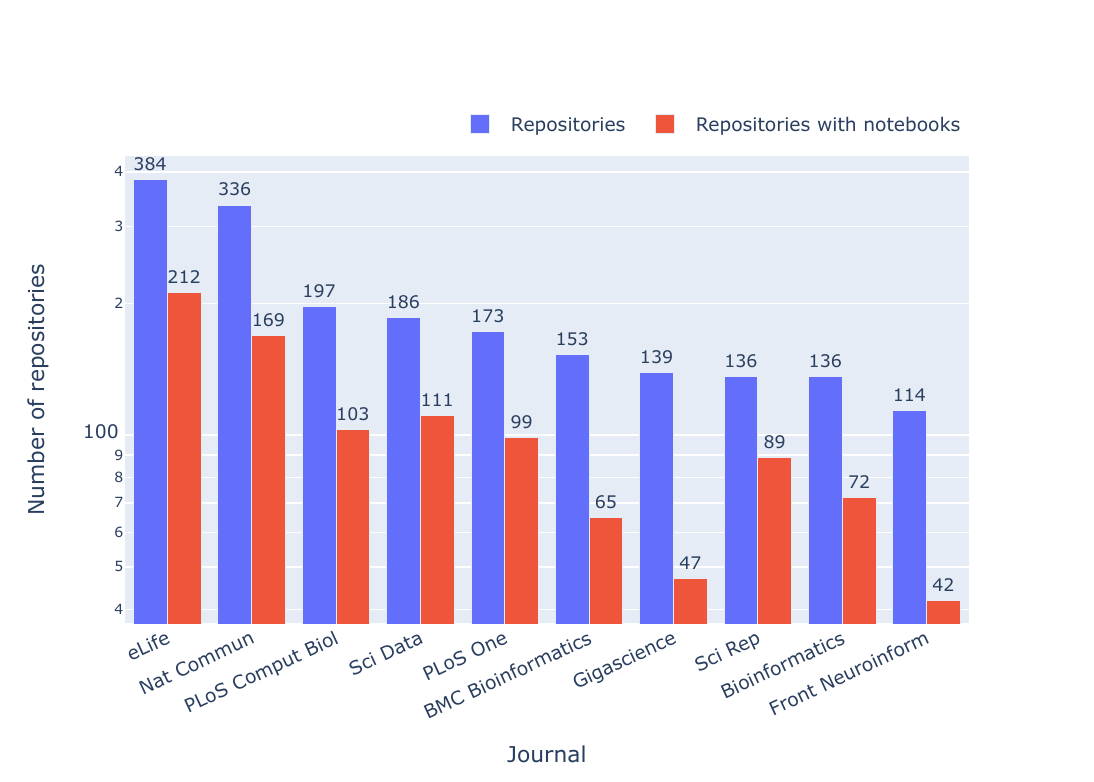}
\caption{
Journals by the number of GitHub repositories and by the number of GitHub repositories with at least one Jupyter notebook.}
\label{fig:Figure_top_journals_repositories_with_without_notebooks}
\end{figure}

\subsection{Journals}
\label{sec:Journals}

In a similar fashion, journals can be ranked by the number of articles that had a valid GitHub repository with at least one Jupyter notebook (cf.\ Figure \ref{fig:Figure_top_journals_with_articles})
or 
by the number of GitHub repositories with and without Jupyter notebooks (cf.\ Figure \ref{fig:Figure_top_journals_repositories_with_without_notebooks}).


The journals \textit{Nature Communications} and \textit{eLife} topped the list in both cases, followed by \textit{PLOS Computational Biology}.
The ratio of GitHub repositories just mentioned to GitHub repositories containing Jupyter notebooks
varies across journals by about a factor of two, with the range being between 
3:1 in \textit{GigaScience}, 2:1 in \textit{Nature Communications},
and 1.5:1 in \textit{Scientific Reports}.
From the 2660 (\initial{1117}) repositories with Jupyter notebooks, 692 (26\%) (\initial{290 (25.9\%)}) 
had one Jupyter notebook, 1,082 (40.7) (\initial{462 (41.4\%)}) had two notebooks, and 618 (23.2\%) (\initial{249 (22.3\%)}) had ten or more notebooks. 
20,838 (76.4\%) (\initial{6,782 (70.4\%)}) of the notebooks belonged to repositories with ten or more notebooks.

\begin{figure}[!htb]
\includegraphics[width=1.0\linewidth]{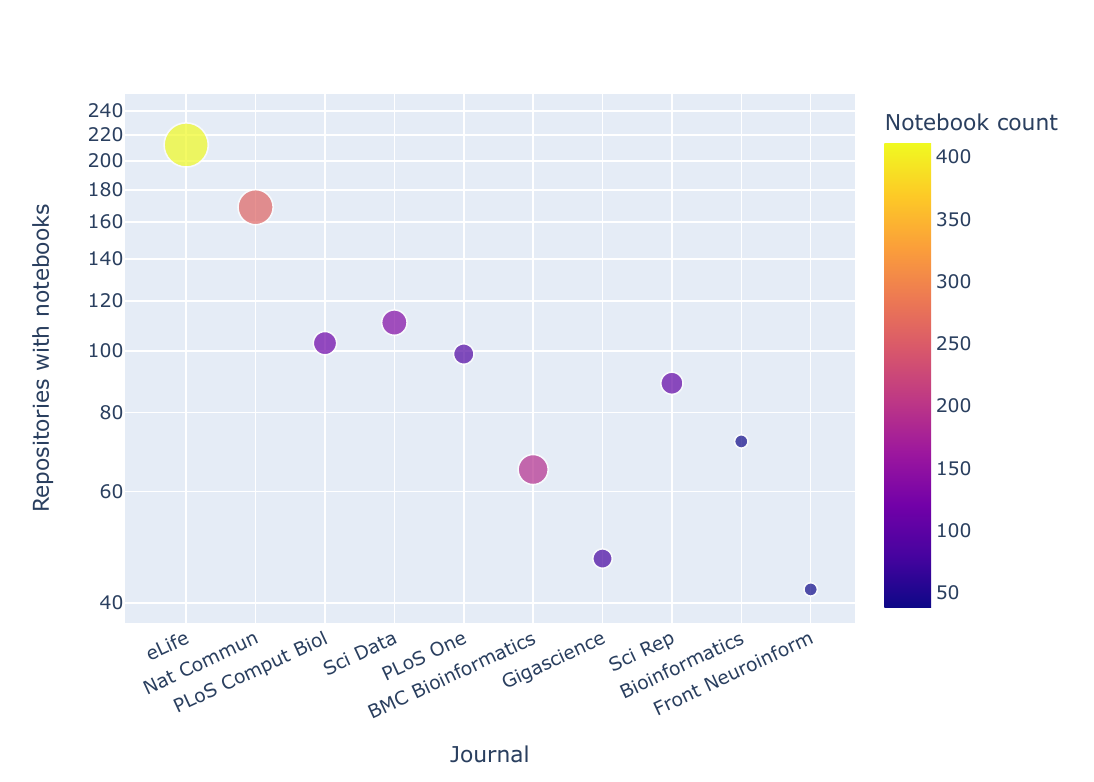}
\caption[Journals by number of GitHub repositories with Jupyter notebooks.]{
Journals by number of GitHub repositories with Jupyter notebooks. 
For each journal, the notebook count gives the maximum number of notebooks within a repository associated with an article published in the journal.}
\label{fig:Figure_top_journals_with_repositories_notebooks}
\end{figure}

\begin{figure}[!htb]
\includegraphics[width=1.0\linewidth]{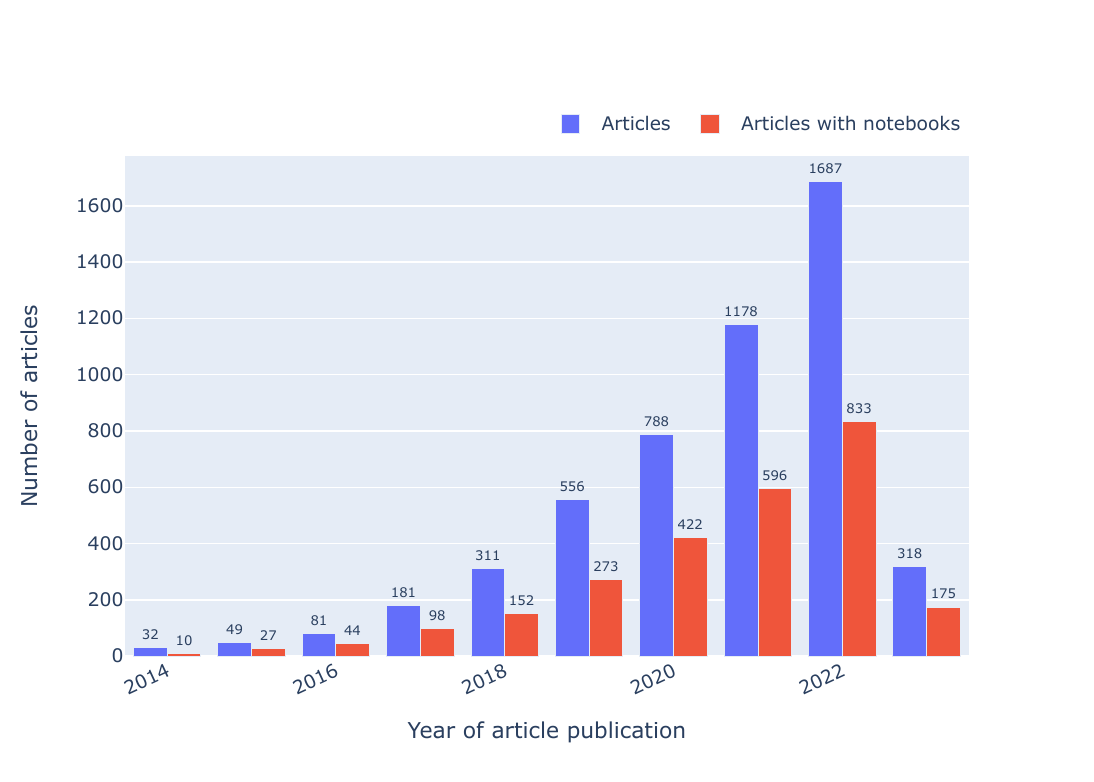}
\caption{
Articles by number of GitHub repositories, highlighting (in red) those with at least one Jupyter notebook, grouped by
year of article publication.
Note that the articles were mined in early 2023, so data for that year are incomplete. However, since we have included the 2023 data in all the non-timeline plots, we decided to keep them in timelines too.
}
\label{fig:Figure_timeline_articles_with_without_notebooks}
\end{figure}
Among the top ten journals with notebooks, \textit{eLife} emerged as the leading journal in terms of the highest number of notebooks, as depicted in Figure \ref{fig:Figure_top_journals_with_repositories_notebooks}. 
Moreover, it ranked first in overall representation when considering journals with repositories containing notebooks.
The growing trend of articles accompanied by Jupyter notebooks is illustrated in Figure \ref{fig:Figure_timeline_articles_with_without_notebooks}, which 
groups articles by year and by number of GitHub repositories containing at least one Jupyter notebook.

\subsection{Programming languages}
\label{sec:ProgrammingLanguages}
\begin{figure}[!htb]
\includegraphics[width=1.0\linewidth]{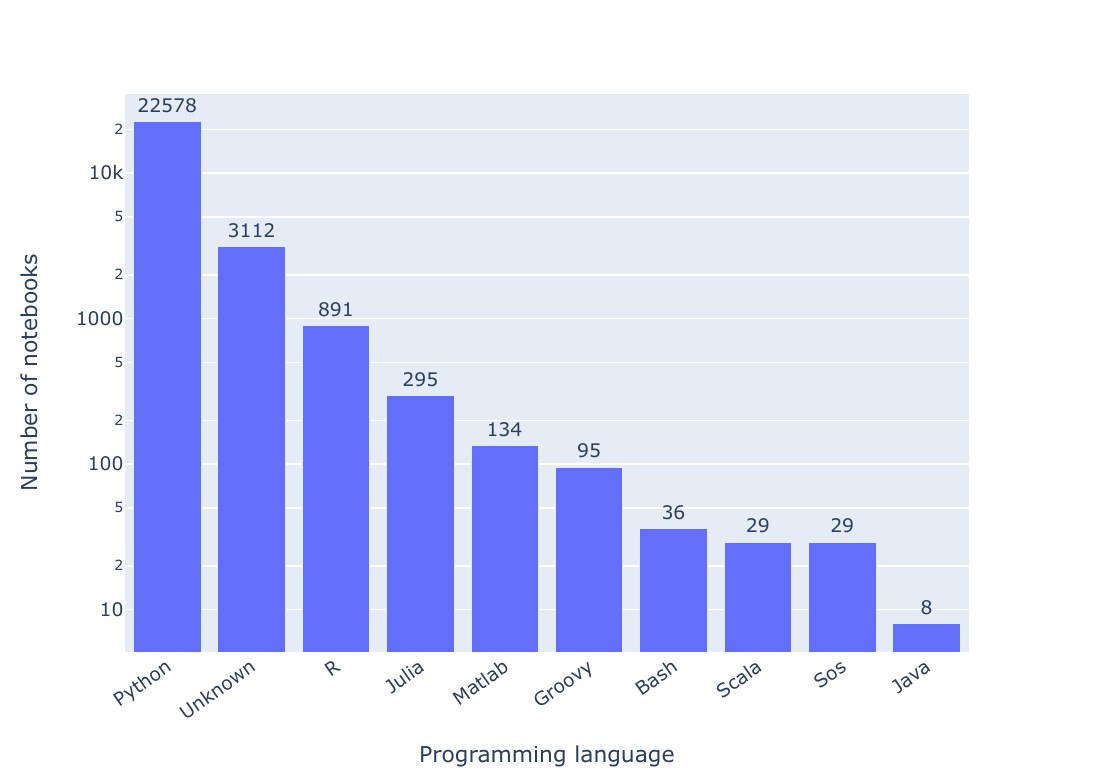}
\caption[Programming languages of the notebooks.]{
Programming languages of the notebooks. ``Unknown'' means the language kernel used was not indicated in a standard fashion. 
}
\label{fig:Figure_top_notebook_language}
\end{figure}

\begin{figure}[!htb]
\includegraphics[width=1.0\linewidth]{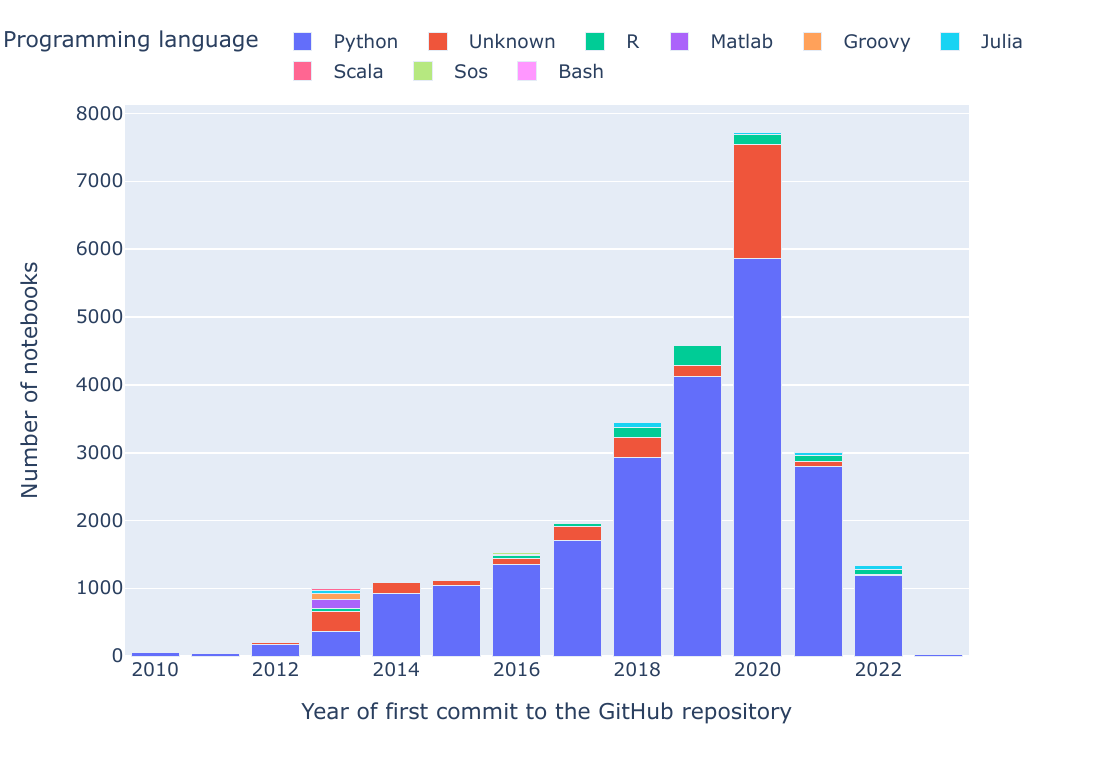}
\caption{
Relative proportion of the most frequent programming languages used  in the notebooks per year.
This analysis includes only programming languages with more than 7 notebooks. In 2023, we observed only 21 Python notebooks, and 
no other programming languages had more than 7 notebooks.
}
\label{fig:Figure_timeline_notebook_language_year}
\end{figure}

The breakdown of the Jupyter notebooks in our corpus by programming language 
(cf.\ Figure \ref{fig:Figure_top_notebook_language}
and
Figure \ref{fig:Figure_timeline_notebook_language_year}
)
shows the three languages behind the Jupyter acronym (\textbf{Ju}lia, \textbf{Pyt}hon and \textbf{R}) 
at the top.
Figure \ref{fig:Figure_top_notebook_language} presents (using a log scale) the most common programming languages used in the notebooks.
Python (82.8\%) (\initial{(84.8\%)}) is the most common programming language, followed by unknown (11.4\%) (\initial{(7.5\%)}), R (3.3\%) (\initial{(4.8\%)}) and Julia (1.1\%) (\initial{0.6\%}).
Unknown notebooks are those which do not declare the programming language or its version in a standard fashion, which is primarily due to early notebooks in which Python was hardcoded, or the language stated in some other non-standard fashion.
A total of 3,112 (\initial{720}) notebooks do not declare a programming language.

There is a steadily increasing use of Python in Jupyter notebooks over the years, as depicted in Figure \ref{fig:Figure_timeline_notebook_language_year}, which presents the top programming languages employed in notebooks based on the year when the article was published.
However, 
the 
rate of change between the initial run 
and the re-run 
differs considerably between languages, as detailed in Table
\ref{tab:language-change}: while some (notably Matlab and Julia) showed marked increases (albeit at low absolute numbers relative to Python), 
others (Groovy, Scala and Java) showed no change, i.e.\ no notebooks from articles published after February 2021.

\begin{table}[!htb]
\caption{Notebook languages from 
Figure \ref{fig:Figure_top_notebook_language}
sorted by the ratio of their frequency in the re-run versus 
in the initial run (cf.\ Figure 7 in \citep{samuel2022computational}).}
\label{tab:language-change}
\begin{tabular}{| L{0.2\linewidth} | R{0.15\linewidth} | R{0.15\linewidth} | R{0.21\linewidth} |}
\toprule
\textbf{Notebook language} & \mbox{} \newline \textbf{re-run} &  \mbox{} \newline \textbf{initial run} & \textbf{ratio \newline re-run/initial} \\
\midrule
Matlab  & 134    & 9           & 14.9                   \\
Julia   & 295    & 59          & 5.0                    \\
Unknown & 3,112   & 720         & 4.3                    \\
Python  & 22,578  & 8,160        & 2.8                    \\
R       & 891    & 461         & 1.9                    \\
Bash    & 36     & 24          & 1.5                    \\
Sos     & 29     & 24          & 1.2                    \\
Groovy  & 95     & 95          & 1.0                    \\
Scala   & 29     & 29          & 1.0                    \\
Java    & 8      & 8           & 1.0                    \\
\bottomrule
\end{tabular}
\end{table}

\subsection{Python versions}
\label{sec:Versions}
\begin{figure}[!htb]
\includegraphics[width=1.0\linewidth]{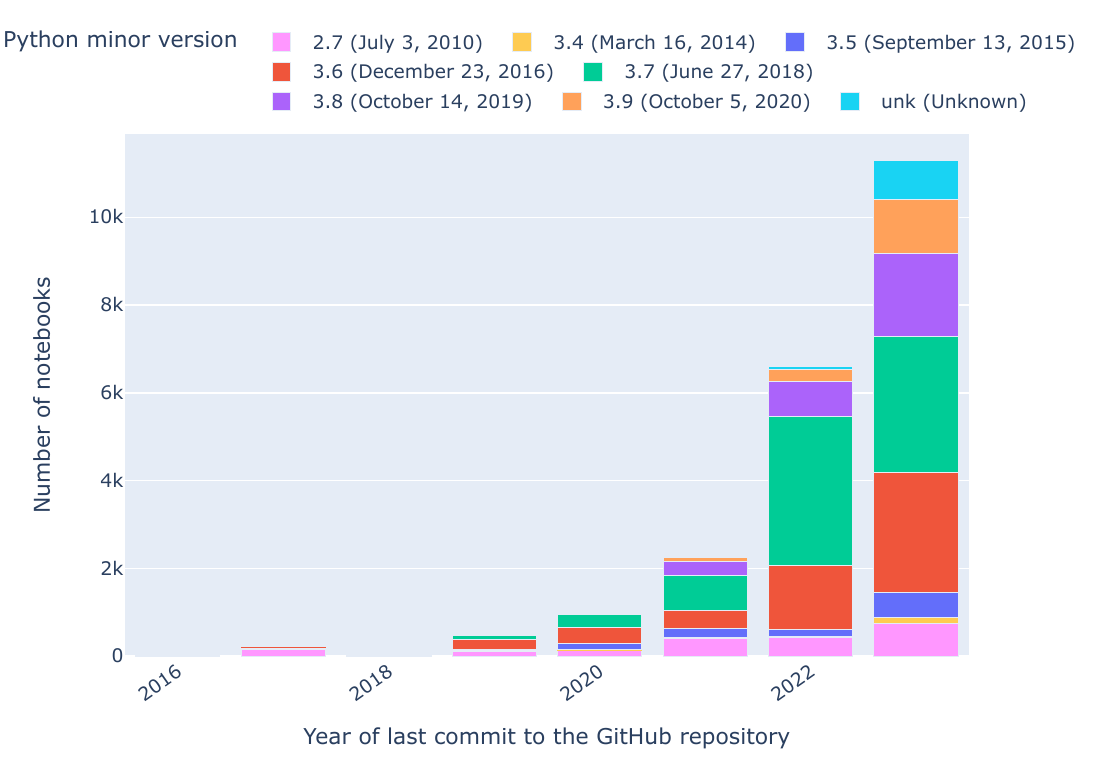}
\caption{
Python notebooks by minor Python version by year of last commit to the GitHub repository containing the notebook. In the legend, the sunset dates for each version are given.
}
\label{fig:Figure_timeline_python_minor_version_by_repo_update}
\end{figure}
\begin{figure}[!htb]
\includegraphics[width=1.0\linewidth]{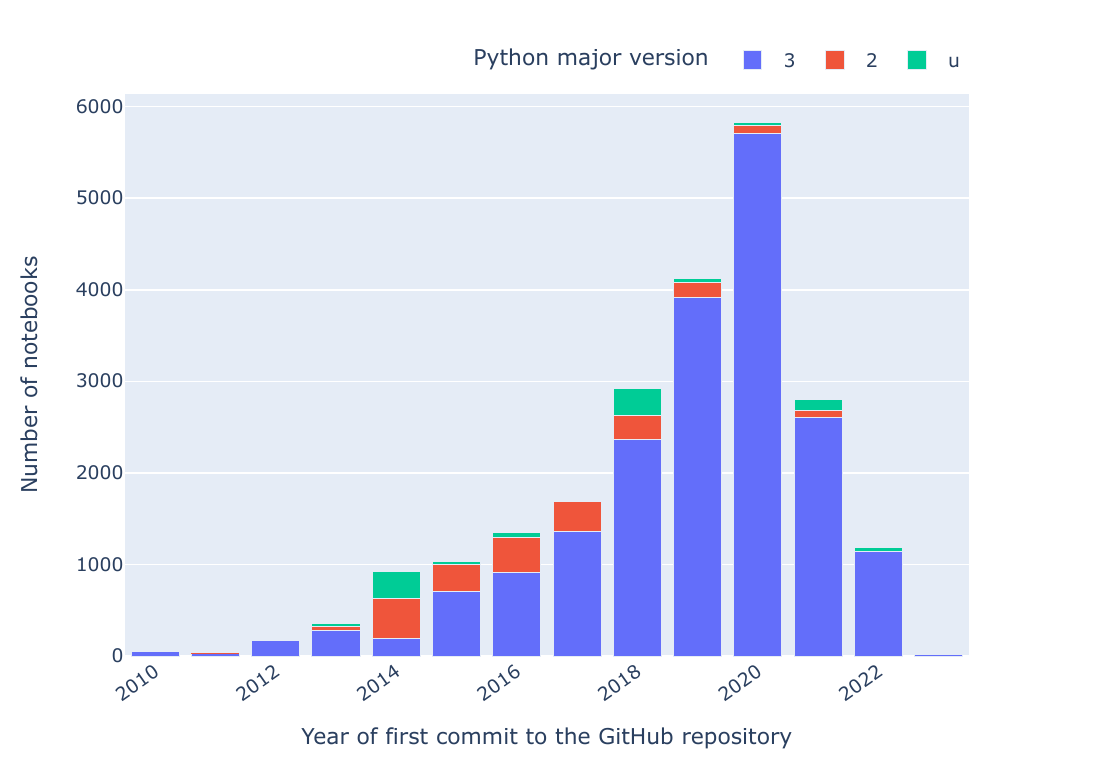}
\caption{
Python notebooks by major Python version by year of first commit to the notebook's GitHub repository.}
\label{fig:Figure_timeline_python_major_version_by_repo_creation}
\end{figure}

When plotting
the Python versions used in notebooks and grouping them by the year in which the repository was last updated (as per Figure \ref{fig:Figure_timeline_python_minor_version_by_repo_update}), it is evident that Python version 3.7 dominates the landscape with 7,667 (\initial{2,031}) notebooks, followed by 5,211 (\initial{2,471}) notebooks with Python version 3.6.
Python version 3.6 and 3.7 are commonly used in recent years, followed by version 3.8 (\initial{2.7}).
There are also some Python notebooks without any version declared.
We see a significant dominance of Python major version 3 in notebooks categorized by the year of the first commit to their GitHub repository (cf.\ Figure \ref{fig:Figure_timeline_python_major_version_by_repo_creation}).
19,508 (\initial{6,028}) notebooks have Python major version 3, 2077 (\initial{1802}) notebooks have Python major version 2, and 954 (\initial{329}) notebooks have an unknown Python version.

\subsection{Notebook structure}
\label{sec:NotebookStructure}
\begin{figure}[!htb]
    \centering
    \begin{subfigure}[t]{0.45\textwidth}
        \centering
        \includegraphics[width=\hsize]{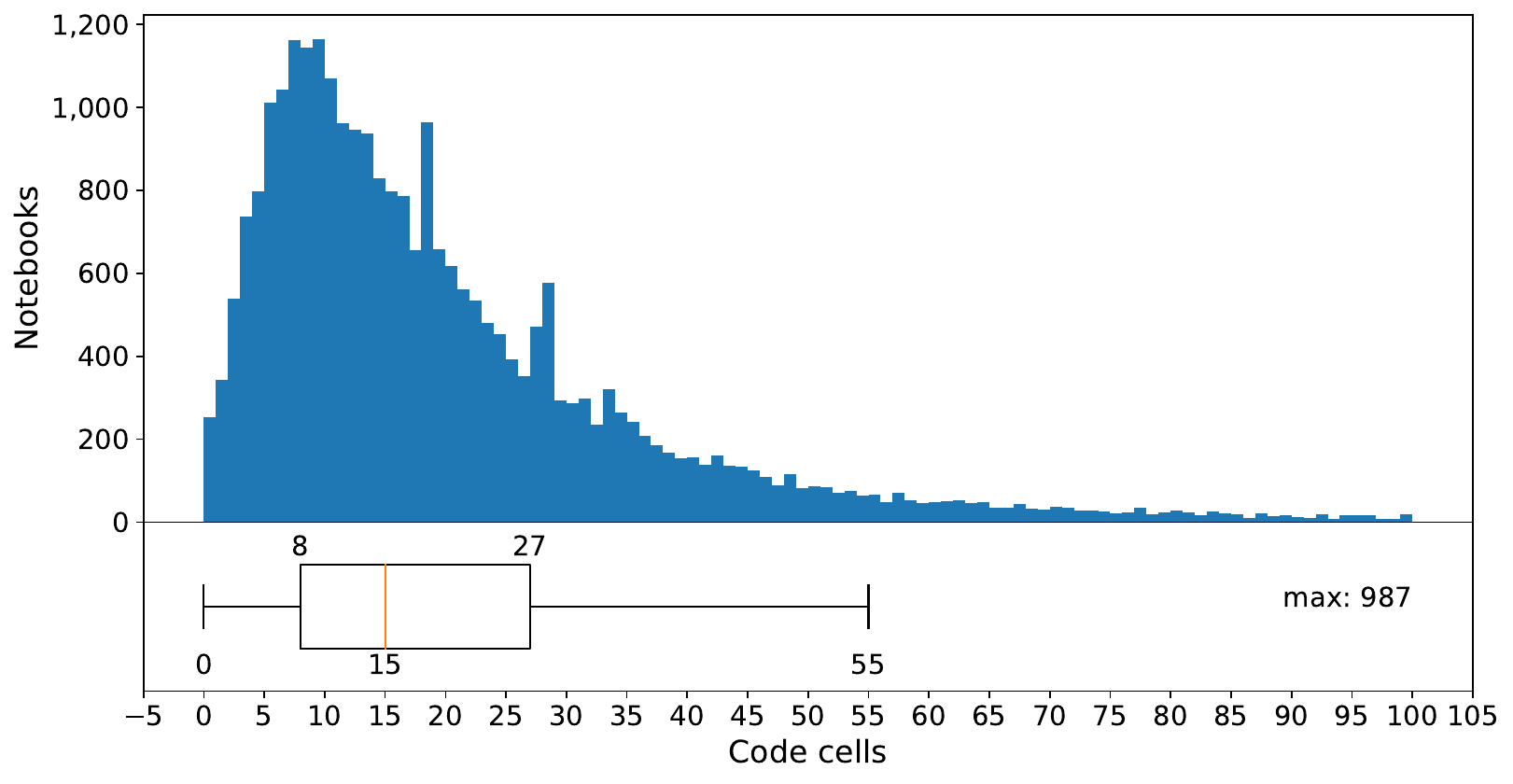}
    \caption{
    Distribution of the number of code cells. }
    \label{fig:Figure_f_code_cells}
    \end{subfigure}
    \hfill
    \begin{subfigure}[t]{0.45\textwidth}
        \centering
        \includegraphics[width=\hsize]{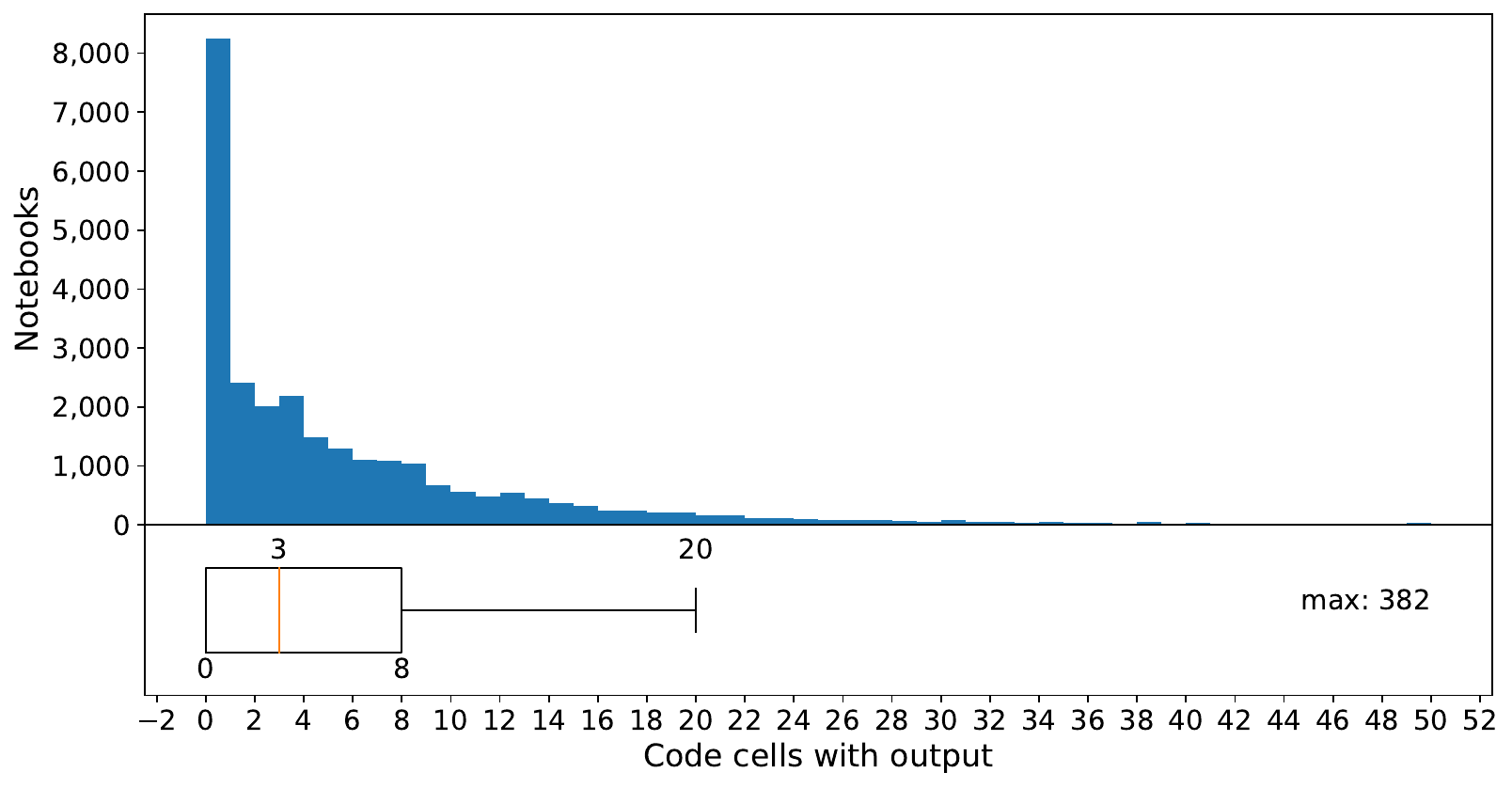}
\caption{
Distribution of the number of code cells with outputs.}
    \label{fig:Figure_f_code_cells_with_output}
    \end{subfigure}

    \vspace{1cm}
    \begin{subfigure}[t]{0.45\textwidth}
        \centering
        \includegraphics[width=\hsize]{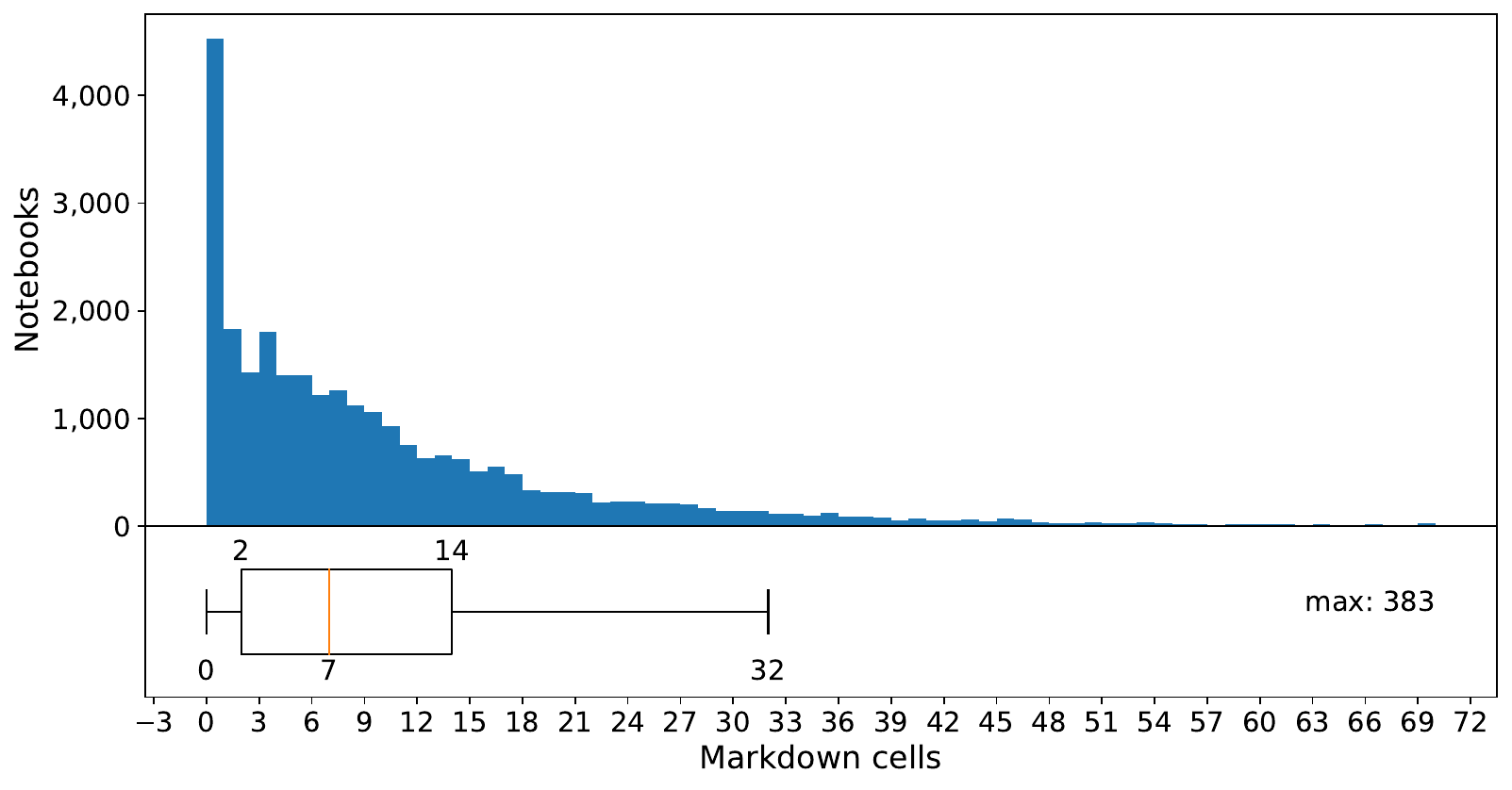}
        \caption{
        Distribution of the number of Markdown cells. }
        \label{fig:Figure_f_markdown}
    \end{subfigure}
    \hfill
    \begin{subfigure}[t]{0.45\textwidth}
        \centering
        \includegraphics[width=\hsize]{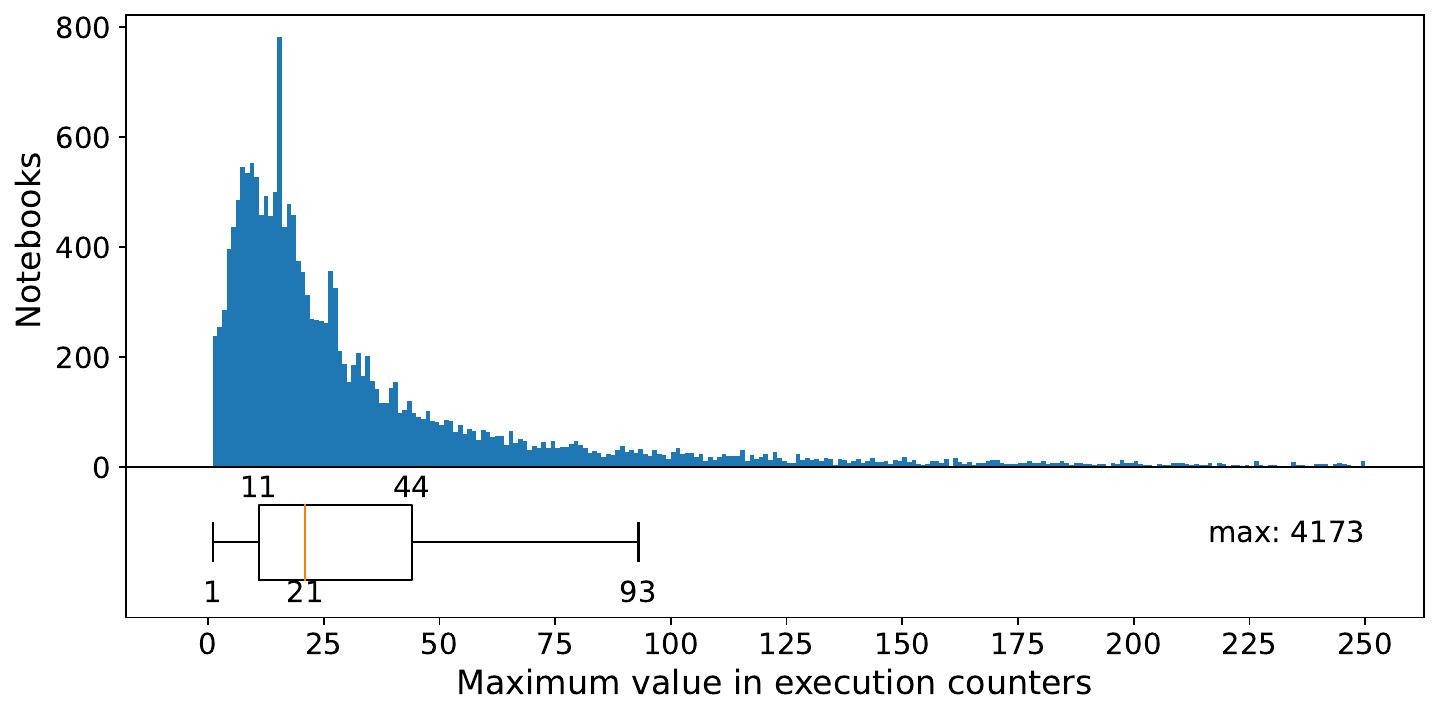}
        \caption{
        Distribution of the maximum execution count.}
        \label{fig:Figure_f_max_execution_count_full}
    \end{subfigure}
    \caption{Analysis of the notebook structure across notebooks in our corpus. The x-axis scale in the diagram depicts the distribution of a particular attribute. The box plot showcases the interquartile range (IQR) along with any outliers beyond the `whiskers'. Annotations highlight values falling below Q1-1.5IQR and above Q3+1.5IQR, serving to identify potential outliers. 
    }
    \label{fig:Figure_notebook_structure_analysis}
\end{figure}

Notebooks have a median of 23 (\initial{20}) cells and 15 (\initial{13}) code cells (Figure \ref{fig:Figure_f_code_cells}).
The average number of cells with outputs in notebooks found in our study is three (\initial{three}), with zero (\initial{zero}) being the least (Figure \ref{fig:Figure_f_code_cells_with_output}).
The maximum number of cells, code cells, and cells with output seen in a notebook are 1,204 (\initial{595}), 987 (\initial{431}), and 382 (\initial{163}), respectively.
The maximum numbers of raw and empty cells seen in a notebook are 57 (\initial{49}) and 160 (\initial{31}), respectively.
Raw cells let the users write output directly, and the kernel does not evaluate them.
The average number of Markdown cells in notebooks is seven (\initial{six}), with the maximum being 383 (\initial{383}) (Figure \ref{fig:Figure_f_markdown}). 
22,733 (83.58\%)
(\initial{6,311 (65.77\%)}) of the notebooks have Markdown cells, while 4,467 (16.42\%)
(\initial{3,284 (34.23\%)}) notebooks do not.
96.35\% (\initial{96.58\%}) of the notebooks use English in the Markdown cells, while 36.77\% (\initial{46.27\%}) notebooks use only English in the Markdown cells.
In addition to English, 
other popular natural languages used in the Markdown are
French (14.09\%) (\initial{11.76\%}) 
and Danish (5.81\%) (\initial{3.96\%}).
In 8,660 (38.09\%) (\initial{1,909 (30.25\%)}) notebooks, we could not detect the language in the Markdown cells.
Further analysis of Markdown cells shows that the average number of lines and words seen in Markdown cells are 24 (\initial{20}) and 127 (\initial{145}), respectively.
Headers and paragraphs, the most commonly seen Markdown elements, appear in 94.69\% (\initial{92.65\%}) and 77.64\% (\initial{81.81\%}) notebooks, respectively.
18,178 (80.62\%) (\initial{6,710 (82.24\%)}) notebooks have execution numbers, while
4,371 (19.38\%) (\initial{1,449 (17.76\%)}) notebooks don't.
The maximum execution count seen in a notebook is 4,173 (\initial{2,076}) (Figure \ref{fig:Figure_f_max_execution_count_full}).

\subsection{Notebook naming}
\label{sec:NotebookNaming}


\begin{figure}[!htb]
\includegraphics[width=1.0\linewidth]{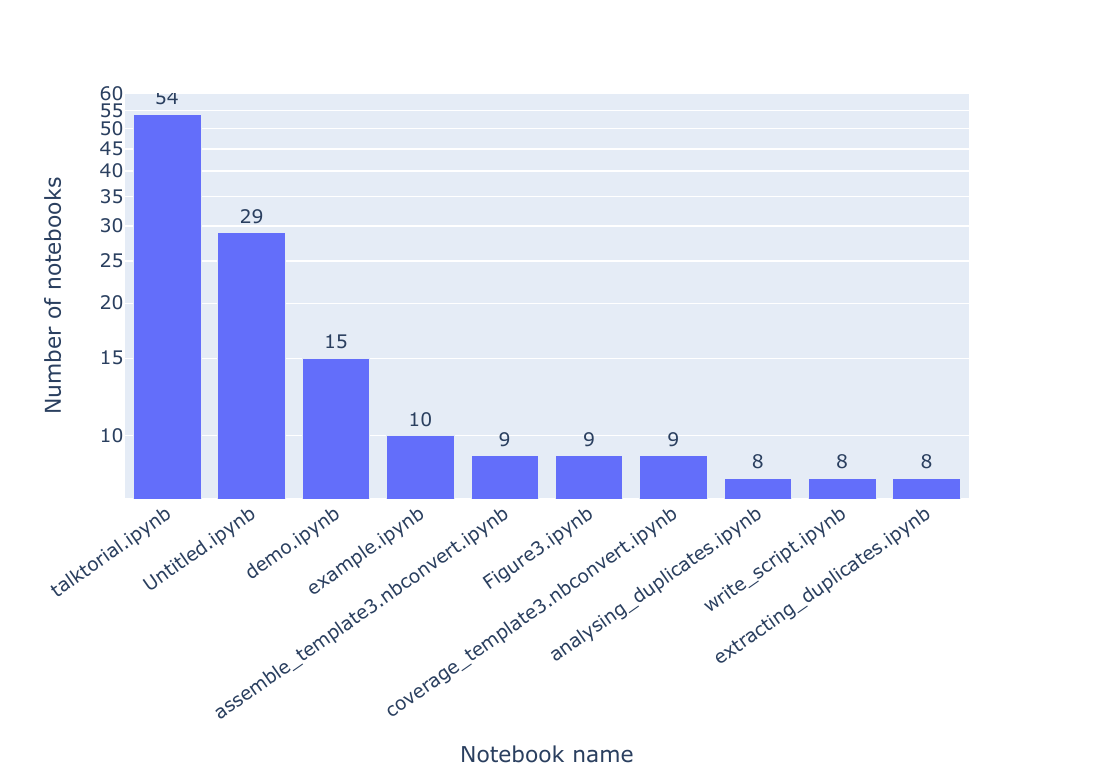}
\caption{
Most frequent notebook titles identified in the re-run results, excluding one repository with hundreds of notebooks whose names would otherwise dominate the list. 
}
\label{fig:Figure_top_notebook_names_excluding_1_repo}
\end{figure}


\begin{figure}[!htb]
\includegraphics[width=1.0\linewidth]{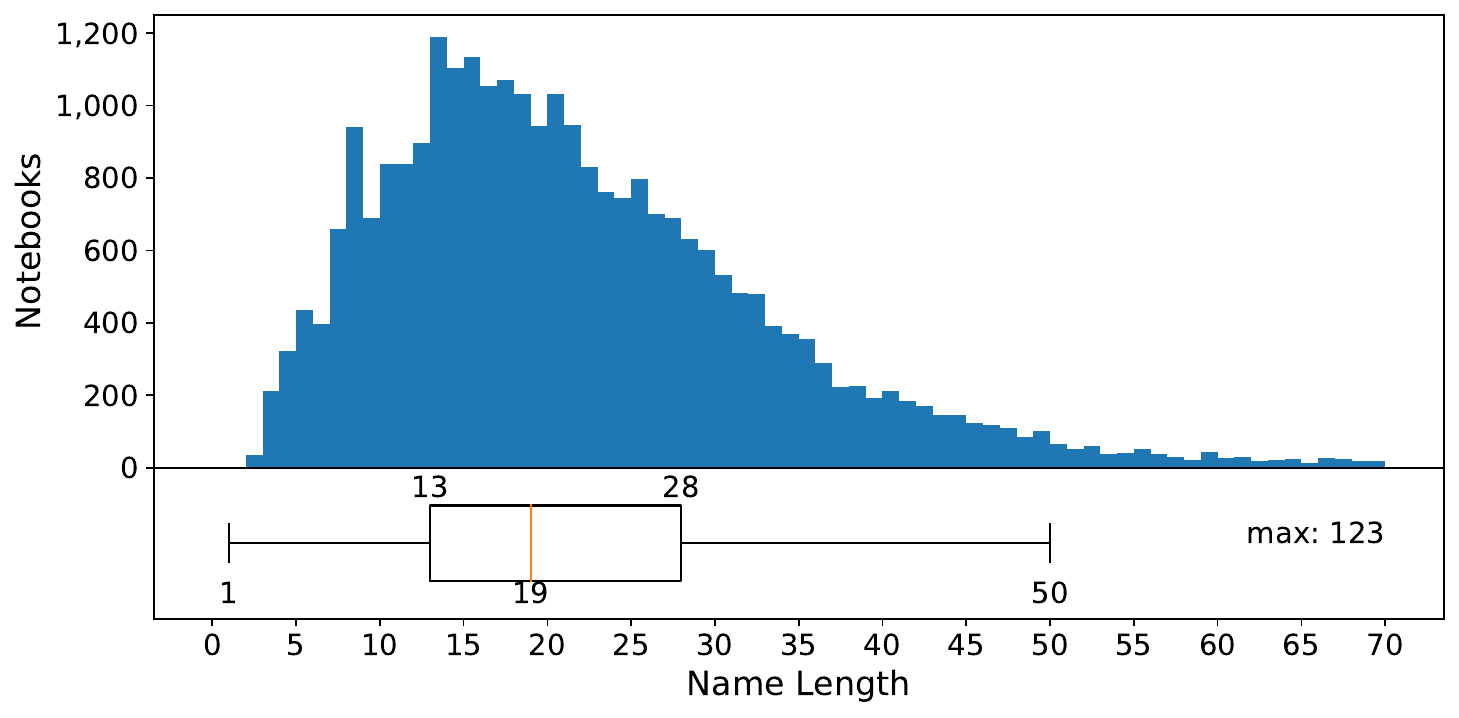}
\caption{
Distribution of notebook title lengths. }
\label{fig:Figure_f_notebook_name_length_full}
\end{figure}

The analysis of notebook titles within our collected data, as depicted in Figure \ref{fig:Figure_top_notebook_names_excluding_1_repo}, reveals the prevalence of certain commonly used names. 
Among them, ``talktorial'', ``Untitled'' and ``demo'' emerged as the top three most frequently encountered notebook titles. 
There are 114 (0.42\%) (\initial{63 (0.65\%)}) notebooks whose title is or starts with ``Untitled'', along with 
68 (0.25\%) (\initial{21 (0.22\%)}) notebooks that contain the name `Copy'.
We also frequently see notebooks with the string `test' in their names.
2,454 (9\%) (\initial{1,070 (11.12\%)}) notebooks have names that are not recommended by the POSIX fully portable filenames guide \citep{pimentel2019a}.
Only 13 (\initial{four}) notebooks have names that are disallowed in Windows.
There are no notebooks without a title (i.e., notebooks with just a '.ipynb' extension).
The average length of the notebook title is 19 (\initial{18}) characters, with a maximum of 123 (\initial{123}) characters and a minimum of 1 
(\initial{2}) (Figure \ref{fig:Figure_f_notebook_name_length_full}).
%
%

\subsection{Notebook dependencies}
\label{sec:NotebookDependencies}

\begin{figure}[!htb]
\includegraphics[width=0.5\textwidth]{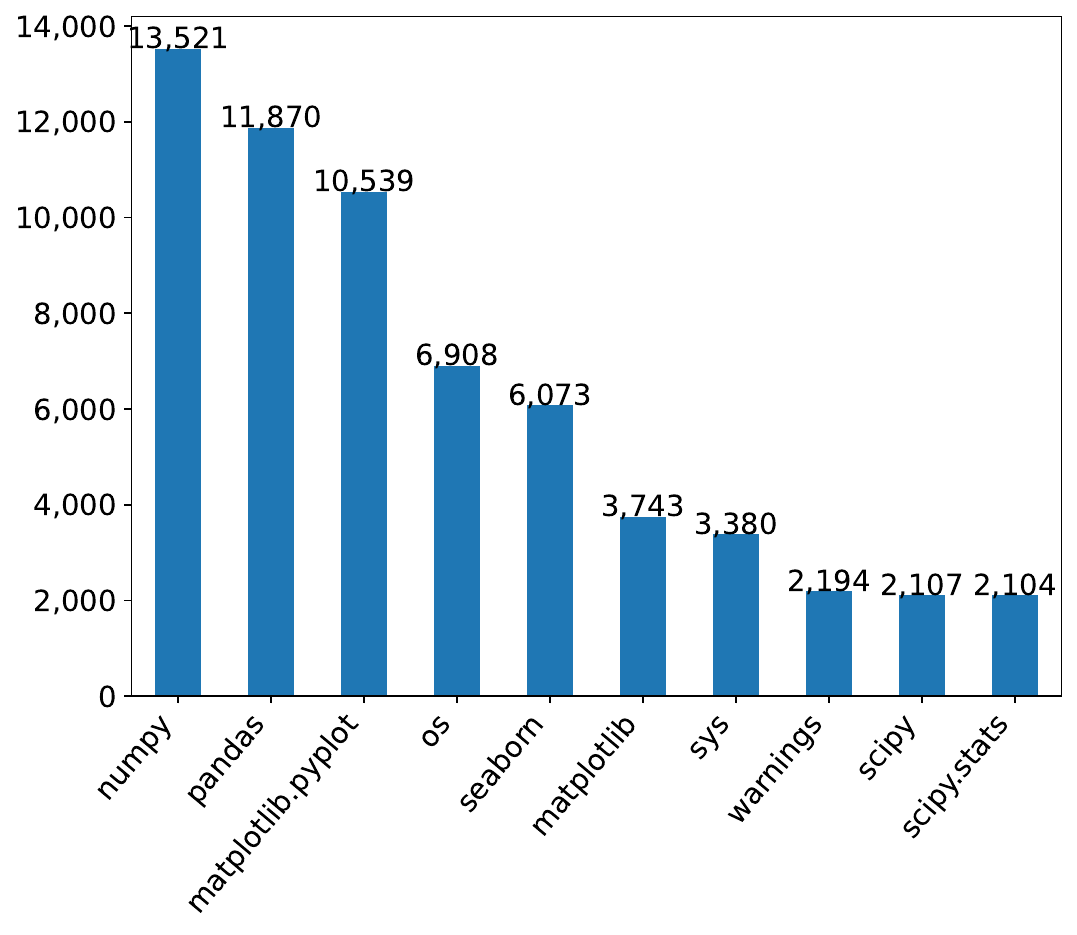}
\caption{
Top Python modules declared in Jupyter notebooks. 
}
\label{fig:Figure_f_notebook_module_full_import}
\end{figure}

\begin{figure}[!htb]
\includegraphics[width=0.5\textwidth]{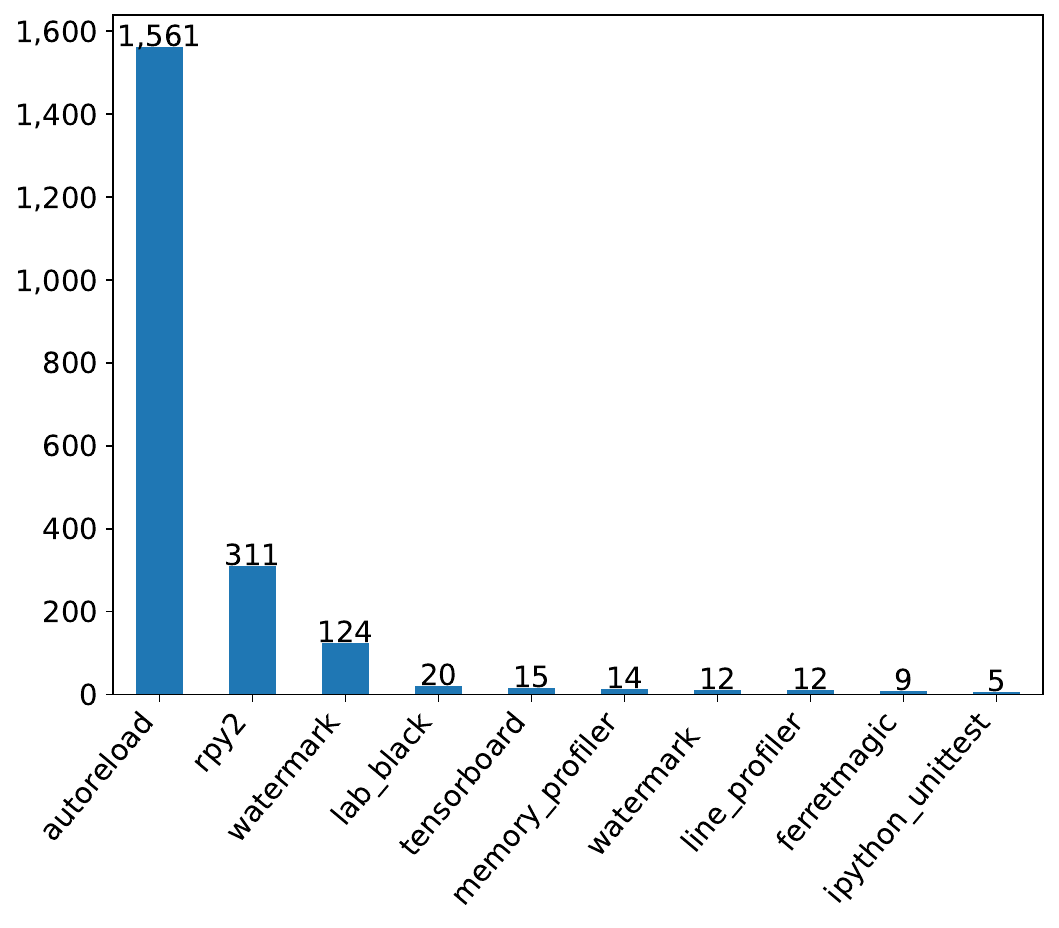}
\caption{
Load extension modules in Jupyter notebooks. 
}
\label{fig:Figure_f_notebook_module_load_ext_toplevel}
\end{figure}

\begin{figure}[!htb]
    \centering
    \begin{subfigure}[t]{0.45\textwidth}
        \centering
        \includegraphics[width=0.75\linewidth]{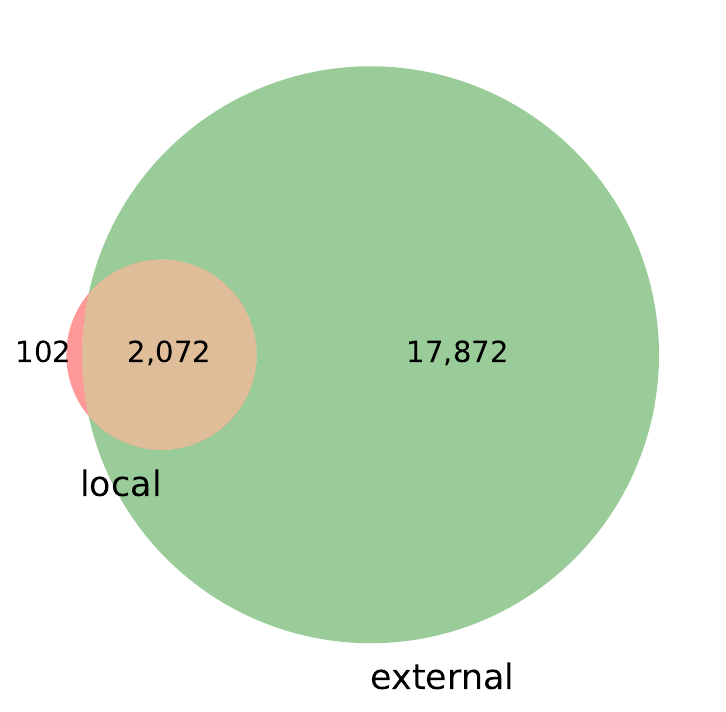} 
        \caption{
        External versus local modules declared in Jupyter notebooks.} \label{fig:Figure_f_notebook_module_external_local}
    \end{subfigure}
    \hfill
    \begin{subfigure}[t]{0.45\textwidth}
        \centering
        \includegraphics[width=\linewidth]{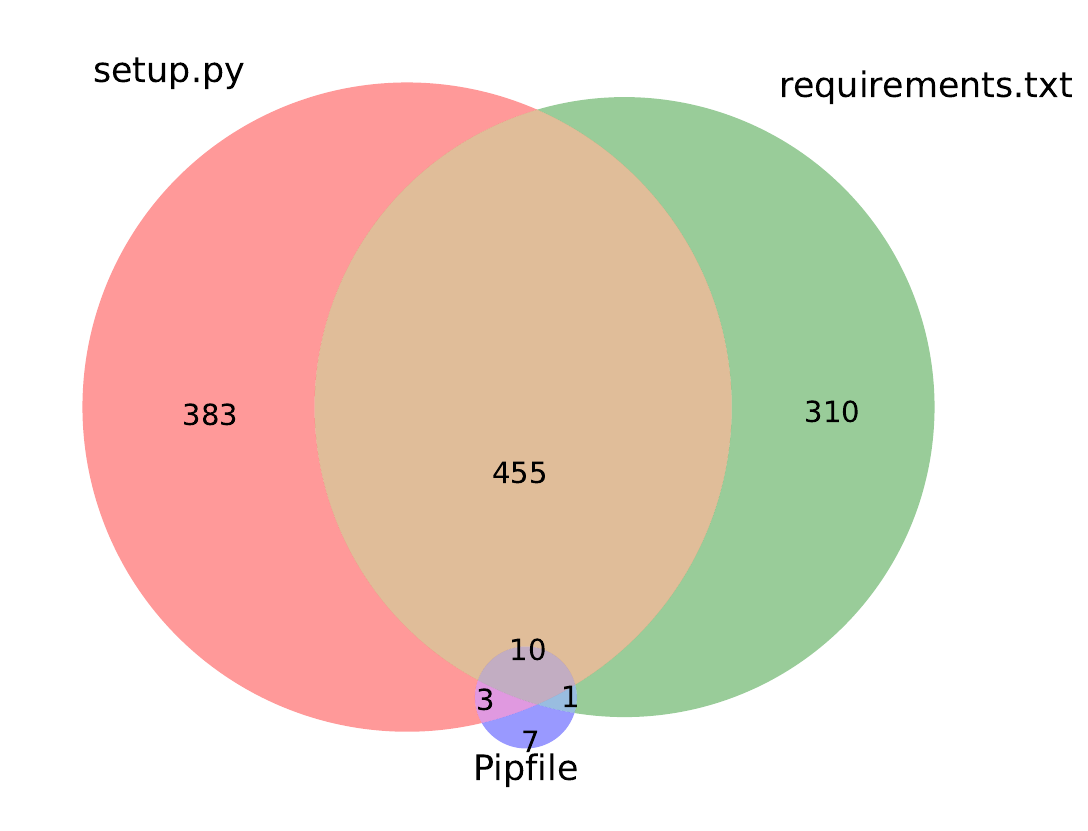} 
        \caption{
        Repositories with dependencies.} 
        \label{fig:Figure_a_repository_dependencies_x3}
    \end{subfigure}

    \vspace{1cm}
    \begin{subfigure}[t]{0.5\textwidth}
    \centering
        \includegraphics[width=0.9\linewidth]{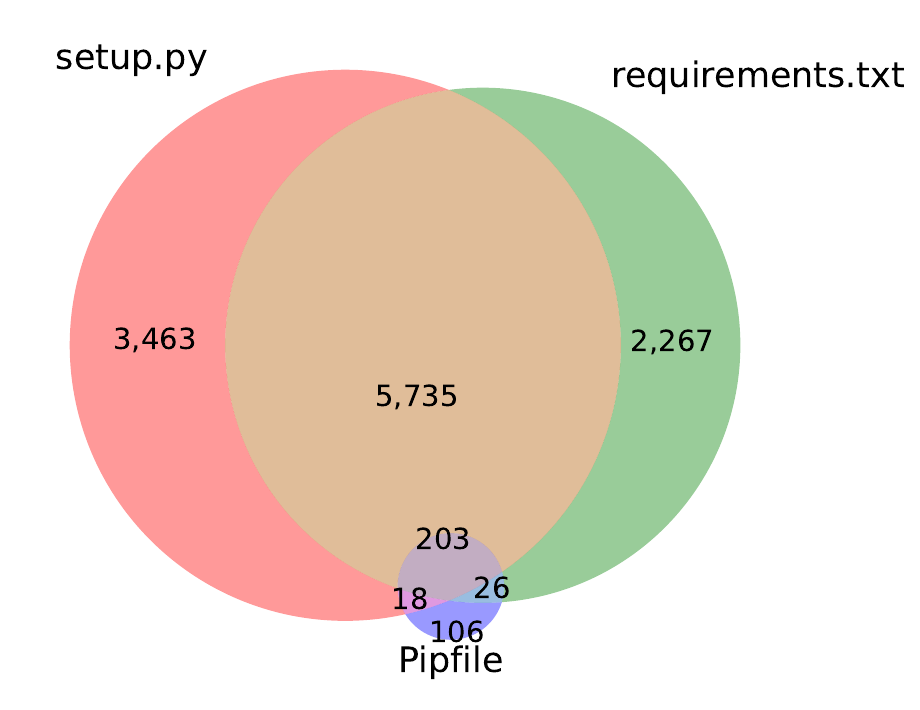} 
        \caption{
        Notebooks with dependencies.} 
        \label{fig:Figure_a_notebook_dependencies_x3}
    \end{subfigure}
    \caption[Dependencies of Juypter Notebooks and GitHub repositories.]{
    Dependencies of Juypter Notebooks and GitHub repositories. In (\subref{fig:Figure_f_notebook_module_external_local}), the notebooks depending on external modules (green) are plotted against notebooks depending on local modules (red) and notebooks that had both (brown) 
    In (\subref{fig:Figure_a_repository_dependencies_x3}) and  (\subref{fig:Figure_a_notebook_dependencies_x3}), GitHub repositories and Jupyter notebooks are shown as to whether they declared their dependencies via any combination of \textit{setup.py} (red),  \textit{requirements.txt} (green) or a \textit{pipfile} (pink).}
    \label{fig:Figure_dependencies}
\end{figure}

Using AST, 
we analyzed the valid Python notebooks.
20,046 (96.85\%) (\initial{5,248 (69.06\%)}) notebooks had imports, of which 2,174 (10.5\%) (\initial{714 (9.40\%)}) had local imports, while 19,944 (96.35\%) (\initial{5,216 (68.64\%)}) had external modules (Figure \ref{fig:Figure_f_notebook_module_external_local}).
Local imports denote the import of modules defined in the notebook repository's directory.
The most used Python modules declared in the notebooks (cf.\ Figure \ref{fig:Figure_f_notebook_module_full_import}) are 
\textit{numpy} (13,521) (\initial{(3,255)}), \textit{pandas} (11,870) (\initial{(2,428)}), and \textit{matplotlib.pyplot} (10,539) (\initial{(2,411)})~--
all widely used
for data manipulation, analytics, and visualizations.

A particular type of software 
used in Jupyter notebooks
are load extensions that provide additional functionality for interacting with the notebook environment. Within our corpus, 
the
most popular ones (cf.\ Figure
\ref{fig:Figure_f_notebook_module_load_ext_toplevel})
included two that are directly related to reproducibility~--
\textit{autoreload} (which reloads modules before executing code that depends on them)
and 
\textit{watermark} (which saves metadata about the environment in which a notebook was run).
Another popular load extension was
\textit{rpy2}, which facilitates the use of \textit{R} code within notebooks running on a Python kernel.


11,818 (\initial{4,650 (48.31\%)}) of the notebooks 
belong to repositories that have declared dependencies using \textit{setup.py}, \textit{requirements.txt}, or \textit{pipfile} (Figure \ref{fig:Figure_a_notebook_dependencies_x3}).
There are 1,169 (\initial{492}) repositories with declared dependencies (Figure \ref{fig:Figure_a_repository_dependencies_x3}).
There are 386 (\initial{194}) repositories with \textit{setup.py} file, 311 (\initial{117}) repositories with \textit{requirements.txt} file. 465 (\initial{180}) repositories have both \textit{setup.py} and \textit{requirements.txt} file.
Only 21 (\initial{10}) repositories are with \textit{pipfile} (0.79\%) (\initial{(0.90\%)}).
In our study, 9,419 (34.54\%) (\initial{3,845 (39.95\%)}) of notebooks use a \textit{setup.py} file, 8,231 (30.18\%) (\initial{2,765 (28.73\%)}) notebooks use \textit{requirements.txt} and only 353 (1.29\%) (\initial{186 (1.93\%)}) notebooks use \textit{pipfile}.

\subsection{Notebook reproducibility}
\label{sec:NotebookReproducibility}
In our reproducibility study, we executed 15,817 (58.15\%) (\initial{4,169 (43.45\%)}) Python notebooks.
The dependencies of the notebooks, as 
declared
in their respective repositories, were installed in conda environments.
However, dependencies of 5,429 (34.32\%) (\initial{1,485 (35.62\%)}) notebooks failed to install.
None of these files were malformed with wrong syntax or conflicting dependencies.
We did not find any missing files that required other requirement files which were unavailable or files that needed external tools.
Hence, the reason for the failed installed error is unknown, and we suspect that it may be related to higher-order dependencies (i.e.\ dependencies of the declared dependencies).
We attempted to execute 10,388 (65.68\%) (\initial{2,684 (64.38\%)}) notebooks for the reproducibility study after successfully installing all the requirements.
However, many notebooks failed to execute 
even after installing all the requirements successfully. 

\begin{figure}[!htb]
\includegraphics[width=1.0\linewidth]{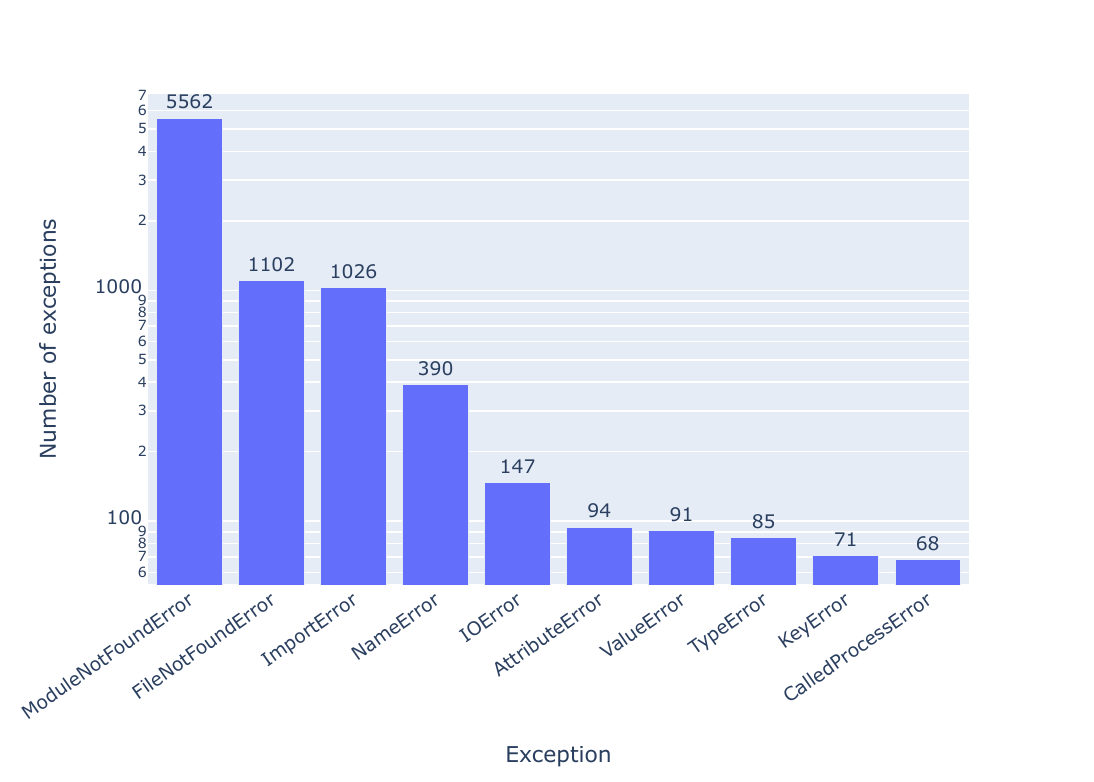}
\caption{
Exceptions occurring in Jupyter notebooks in our corpus. See Table \ref{tab:error-fixes} for information about the nature of these errors and potential fixes.
}
\label{fig:Figure_top_exception_by_reason}
\end{figure}

\begin{figure}[!htb]
\includegraphics[width=1.0\linewidth]{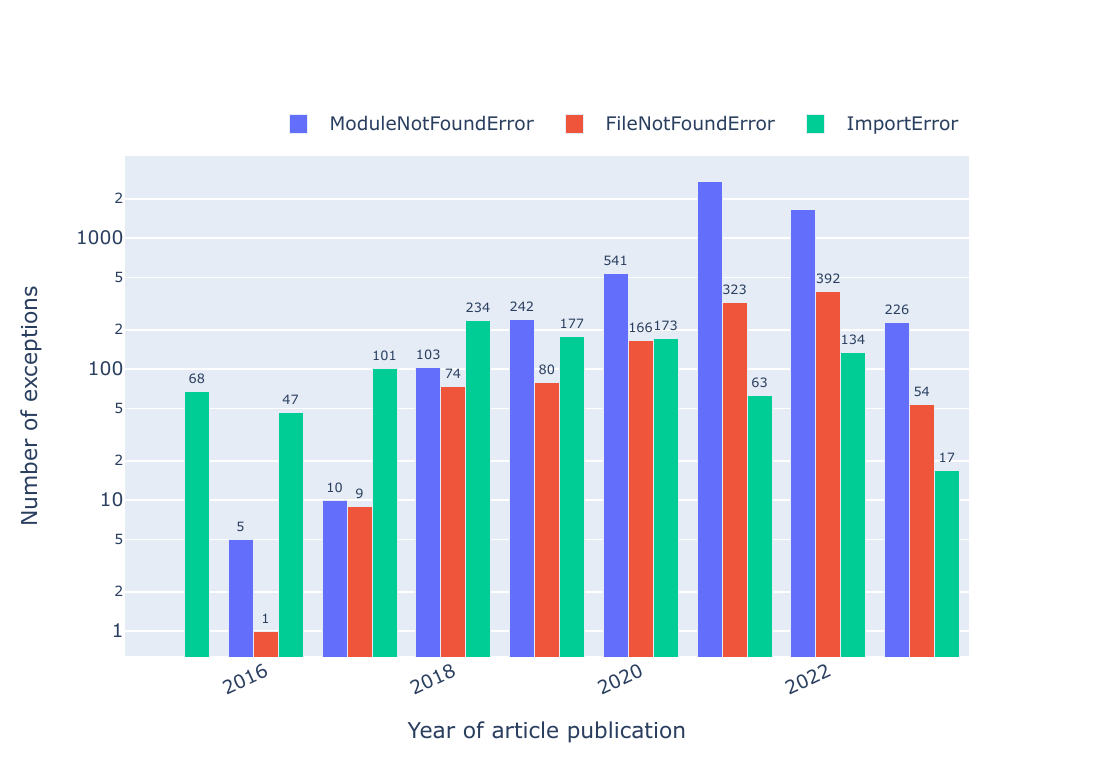}
\caption{
\emph{ModuleNotFoundError}, \emph{ImportError}, and \emph{FileNotFoundError} exceptions by year of publication. Note that data for 2023 is incomplete.}
\label{fig:Figure_timeline_exceptions_per_article}
\end{figure}

\subsection{Exceptions}
\label{sec:Exceptions}


As shown in Figure \ref{fig:Figure_PRISMA}, 
9,100 (87.6\%) (\initial{2,265 (84.39\%)}) notebooks resulted in exceptions,
for a variety of reasons.
\textit{ModuleNotFoundError}, \textit{FileNotFoundError} and \textit{ImportError} are the most common 
exceptions we observed
in the notebooks (Figure \ref{fig:Figure_top_exception_by_reason}).
6,588 (41.65\%) (\initial{1,362 (32.67\%)}) of the executions failed because of \textit{ModuleNotFoundError} and \textit{ImportError} exceptions.
\textit{ModuleNotFoundError} exception occurs when a Python module used by the notebook could not be found.
\textit{ImportError} exception occurs when a Python module used by the notebook could not be imported.
These two errors occur mainly due to missing dependencies.
390 (2.47\%) (\initial{132 (3.17\%)}) notebooks have \textit{NameError}, which occurs when a declared variable in the notebook is not defined.
1,249 (7.9\%) (\initial{374 (8.97\%)}) notebooks have FileNotFoundError or IOError.
These exceptions occur when absolute paths are used to access data or when the data files are not included in the repository.
Overall, 86.29\% of the notebooks we ran returned exceptions that occurred more than 10 times.

The relationship between the top three common exceptions, namely \textit{ModuleNotFoundError}, \textit{ImportError}, and \textit{FileNotFoundError}, and the publication year of the articles is depicted in Figure \ref{fig:Figure_timeline_exceptions_per_article}
as a function of the year of publication of the associated articles.
%
It shows
an increase in the \textit{ModuleNotFoundError} over the years following its introduction with Python 3.6 in 2016, overtaking \textit{ImportError} by 2019.


\begin{figure}[!htb]
\includegraphics[width=1.0\linewidth]{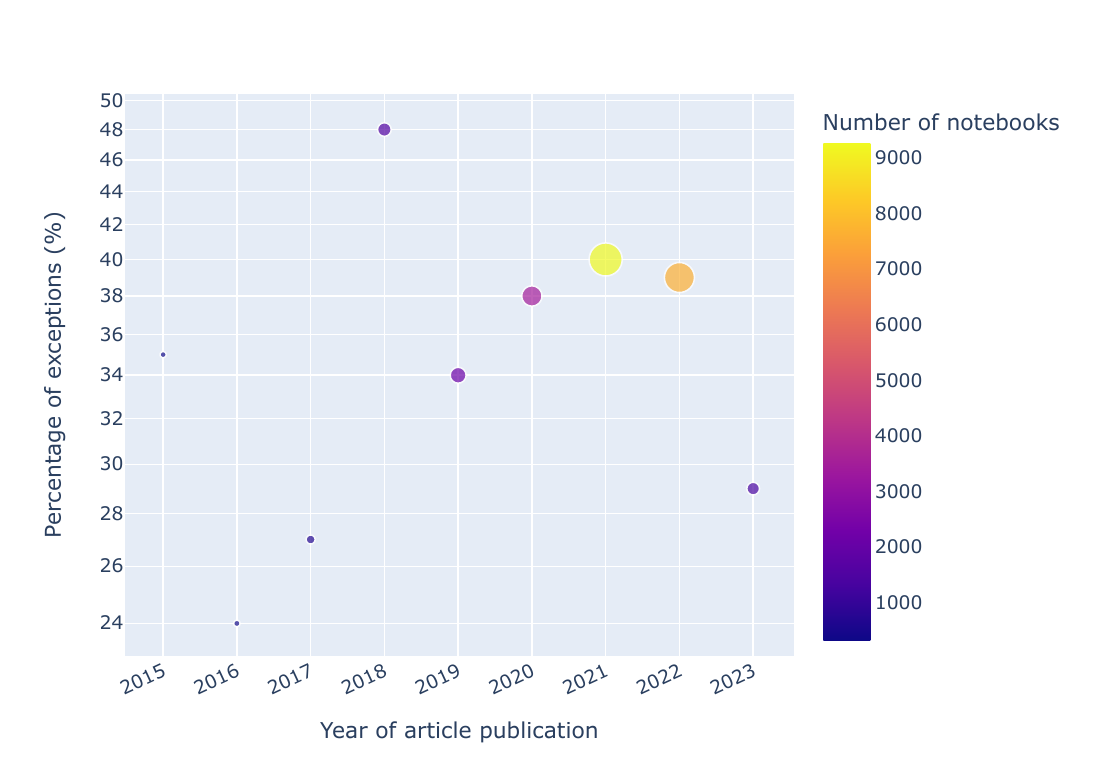}
\caption{
Exceptions by year of publication normalized by the number of notebooks associated with articles published that year.
}
\label{fig:Figure_timeline_exceptions_by_year_notebook_scatter_percentage}
\end{figure}

We observe that the number of exceptions by the year of publication normalized by the number of notebooks peaked in 2021, as shown in Figure \ref{fig:Figure_timeline_exceptions_by_year_notebook_scatter_percentage}. 
If that trend holds, the numbers for 2023 (where data are currently incomplete) would
be expected to be lower than for 2022.
In either case, it would be interesting to explore 
in more detail
the factors that 
contribute to this development.

Apart from such general trends across our entire corpus, we can slice the data in various ways to explore how the frequency of exceptions in Jupyter notebooks relates 
to a range of variables.
For some of these (subsequently \textbf{bolded}), we will briefly outline pertinent observations.

\begin{figure}[!htb]
\includegraphics[width=1.0\linewidth]{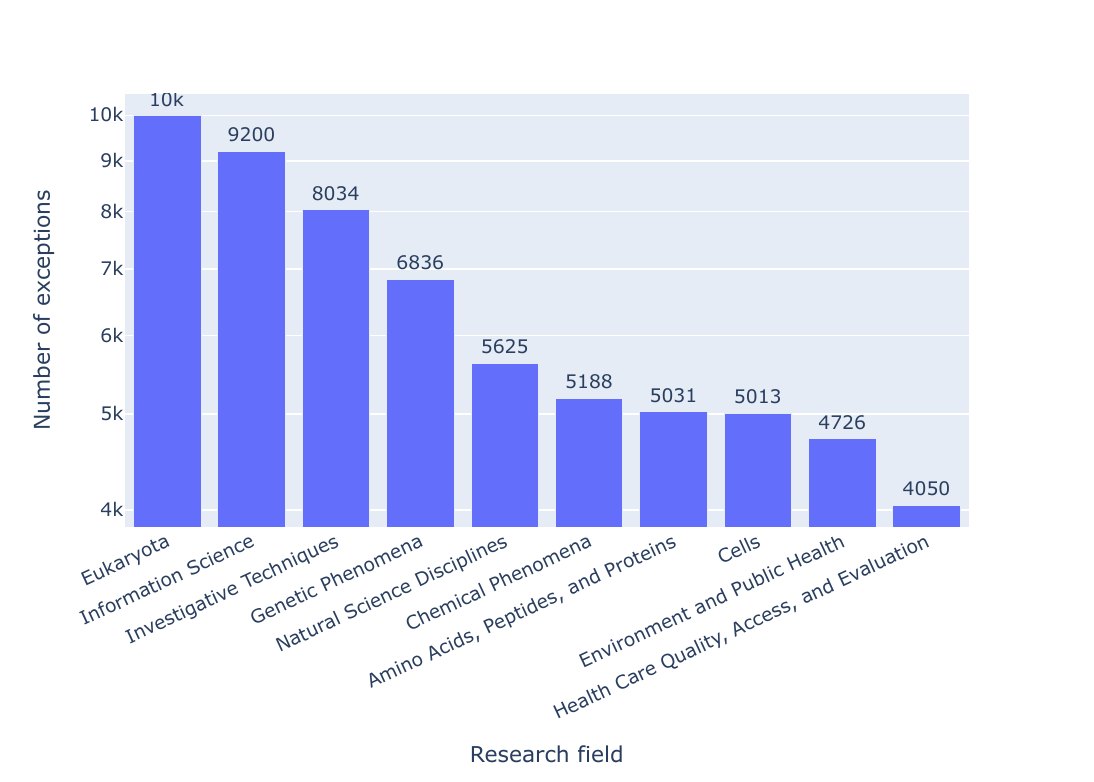}
\caption{
Jupyter notebook exceptions by research field, taking as a proxy the highest-level MeSH terms (which may be more than one) of the article associated with the notebook. We did not normalize these values, so as to let the magnitude of the problem speak for itself.}
\label{fig:Figure_exceptions_by_field}
\end{figure}



\begin{figure}[!htb]
\includegraphics[width=1.0\linewidth]{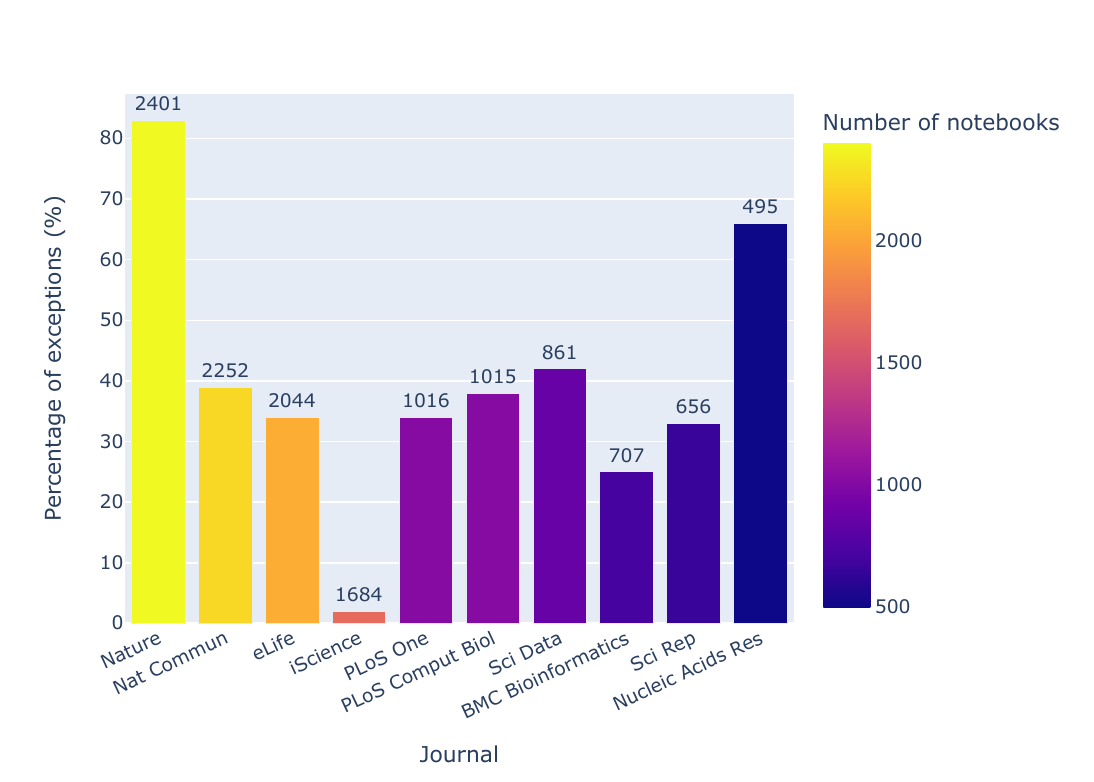}
\caption{
Exceptions by journal, normalized by the number of notebooks and sorted by the notebook count and percentage of exceptions. 
The absolute number of notebooks associated with a journal is presented on top of its bar. As an example, in the journal \textit{iScience}, 26 exceptions were identified among 1684 notebooks, accounting for 2\% of the total.
For context, \textit{Gigascience} had 116 exceptions in 405 notebooks, giving it an exception percentage of 29\%.
}
\label{fig:Figure_exceptions_by_journal_notebooks_by_percentage}
\end{figure}

\begin{figure}[!htb]
\includegraphics[width=1.0\linewidth]{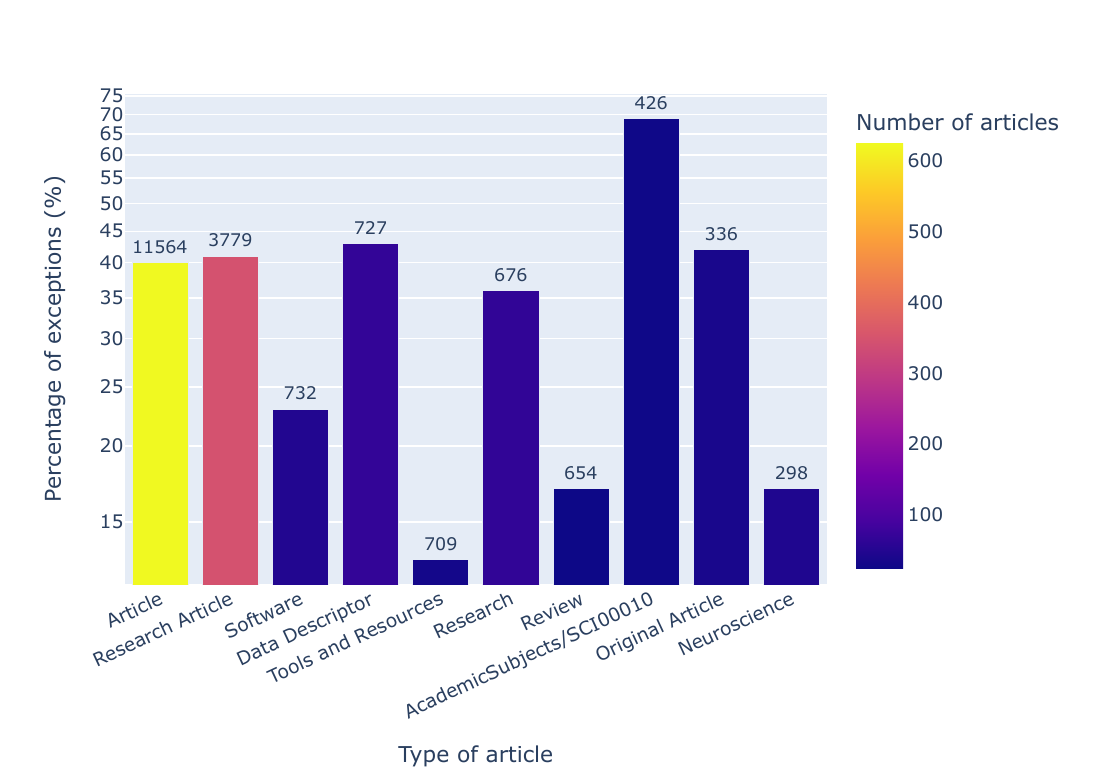}
\caption{
Exceptions by article type, normalized by the number of notebooks per article type and sorted by the total number of notebooks per article type, which
is shown on top of each bar. For example, out of 709 notebooks associated with \textit{Tools and Resources} articles~-- published in \textit{eLife}~\citep{Schekman2015Recognizing}~-- 13\% resulted in exceptions, but there were only 32 such articles in total.
The tag \textit{AcademicSubjects/SCI00010} is used by Oxford University Press to identify articles in biology, for which the exception rate was about five times that of \textit{Tools and Resources} articles.
}
\label{fig:Figure_exceptions_by_subject_article_normalized}
\end{figure}

In terms of \textbf{research field}, we can take MeSH terms 
as a proxy, i.e.\  
we can assign MeSH terms to the articles we mined from PubMed Central, and plot the frequency of exceptions of Jupyter notebooks associated with these articles as a function of those MeSH terms, as per
Figure 
\ref{fig:Figure_exceptions_by_field}. The main finding here 
is that exceptions come in great numbers for any of the areas in which Jupyter is a popular tool.


The relationship between Jupyter notebook exceptions and \textbf{journals} can be explored via Figure \ref{fig:Figure_exceptions_by_journal_notebooks_by_percentage}, which highlights some journals with exception rates well above 50\% (\emph{Nature} and \emph{Nucleic Acids Research}) as well as some well below that mark (\emph{iScience} and \emph{BMC Bioinformatics}), 
indicating better reproducibility.


In a similar fashion, 
%
the distribution of exceptions across different \textbf{types of articles} is illustrated in Figure \ref{fig:Figure_exceptions_by_subject_article_normalized}. 
Notably, technically oriented article types like \emph{Tools and Resources} or \emph{Software} perform better than average, while biological articles in journals published by Oxford University Press (the current publisher of \textit{Gigascience}) underperform in this regard. 
%

\begin{figure}[!htb]
    \centering

    \begin{subfigure}[t]{0.45\textwidth}
        \centering
        \includegraphics[width=\hsize]{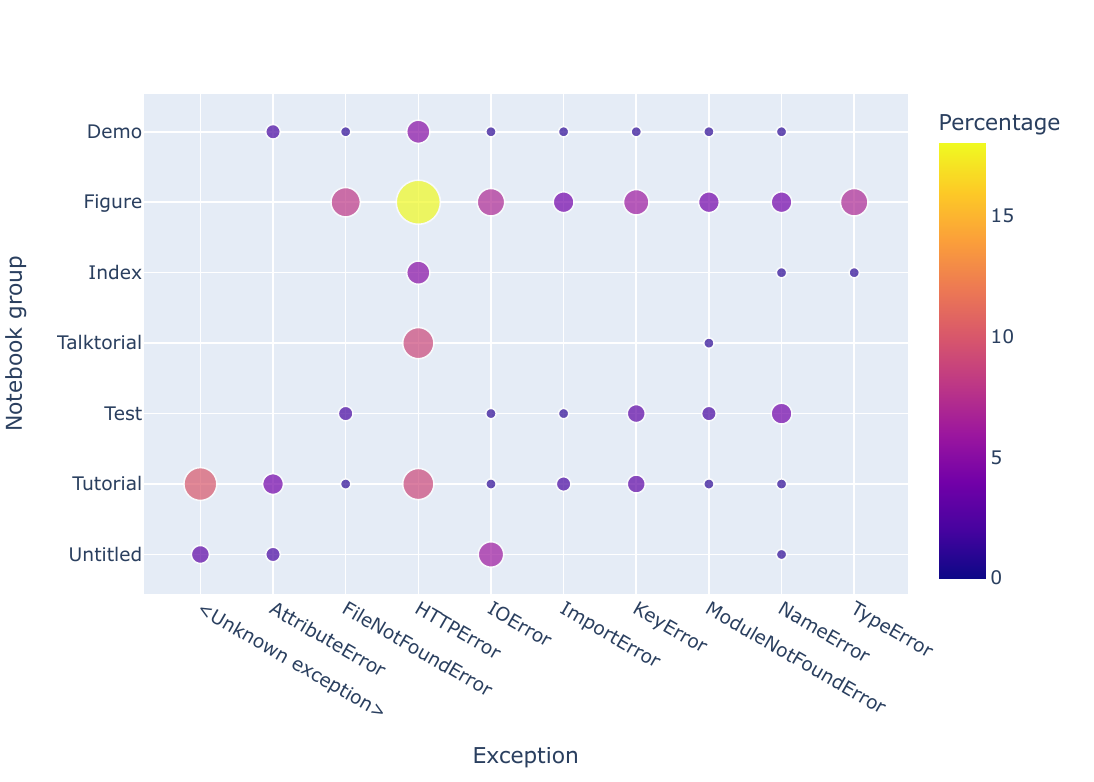}
        \caption{
        Correlation matrix for 
        common name strings
        and exceptions (both in alphabetical order) observed for notebooks. 
        Out of the 10k notebooks in the corpus overall, 847 
        contained any of the 
        (case-normalized)
        target strings 
        (of these, 
        93 had ``demo'', 
        373 ``figure'', 
        14 ``index'', 
        54 ``talktorial'', 
        156 ``test'', 
        119 ``tutorial'', 
        38 ``untitled'').
        For instance, the yellow marker indicates that 18\% (4 out of 22) of the notebooks giving an \textit{HTTPError} had a ``figure'' string in their name.
        }
        \label{fig:Figure_correlation_selected_name_exception}
    \end{subfigure}
    \begin{subfigure}[t]{0.45\textwidth}
        \centering
        \includegraphics[width=\hsize]{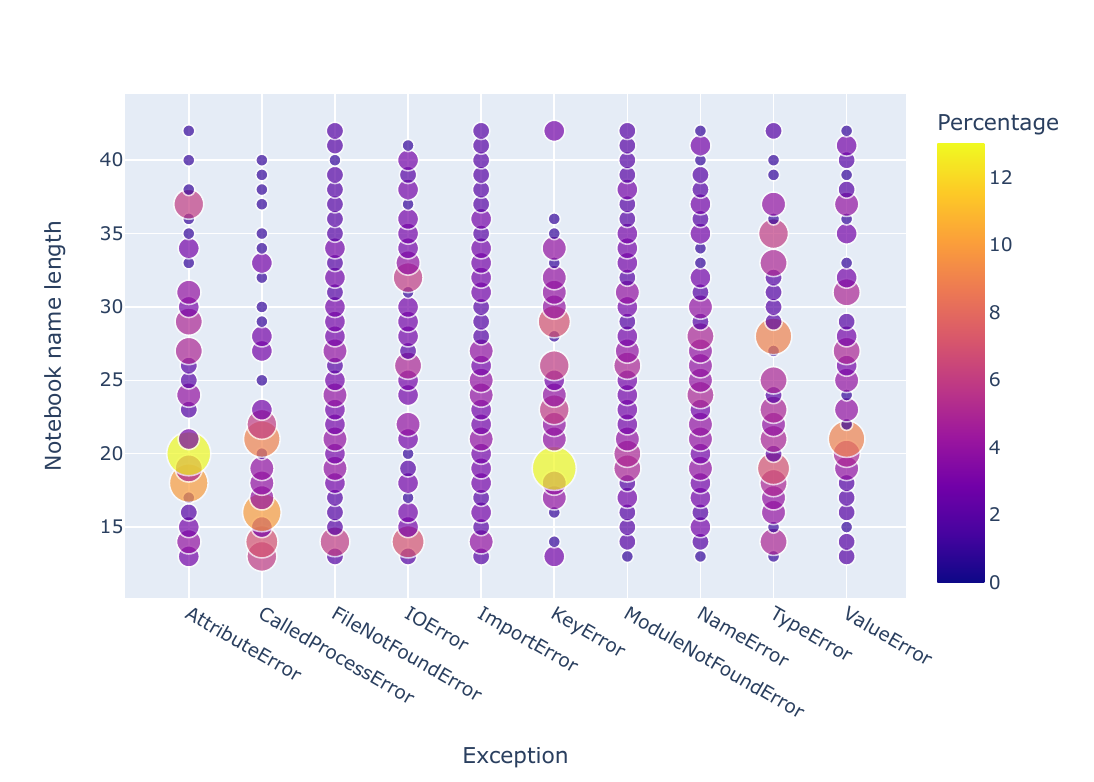}
        \caption{ Correlation matrix for notebook name length and the exceptions raised by the associated notebooks. For instance, 13\% of all \textit{KeyError} exceptions were raised by notebooks with filenames of 19 characters.
        }
        \label{fig:Figure_correlation_namelen_exception}
    \end{subfigure}
    \begin{subfigure}[t]{0.45\textwidth}
        \centering
        \includegraphics[width=\hsize]{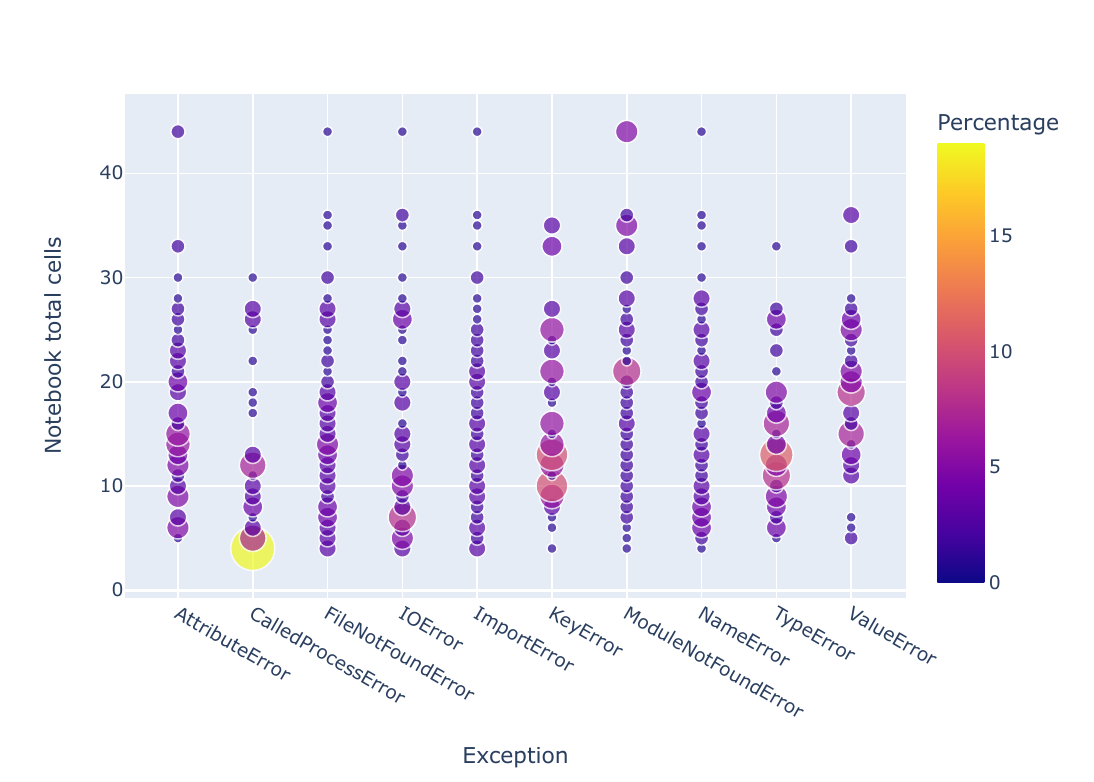}
        \caption{ Correlation matrix for total number of notebook cells and the exceptions raised by the respective notebooks. For example, 
        18\% of all \textit{CalledProcessError} exceptions were 
        due to
        notebooks with 
        4 cells.
        }
        \label{fig:Figure_correlation_totalcells_exception}
    \end{subfigure}
    \caption{Analysis of the notebook structure and exceptions. 
    In all three panels, ``Percentage'' represents the percentage of exceptions from notebooks with a given ordinate value relative to the total number of notebooks 
    with that 
    exception.
    }
    \label{fig:Figure_correlation}
\end{figure}

While exploring correlations between \textbf{notebook file names} and exceptions (cf.\ Figure
\ref{fig:Figure_correlation_selected_name_exception}), 
some patterns 
begin to emerge,
e.g.\ talktorials tend to cause fewer exceptions than notebooks related to figures, while unknown exceptions are frequent in tutorial notebooks.
In terms of \textbf{file name length}, some exception types are more frequent for shorter file names, while other exception types are distributed relatively uniformly across different file name lengths, as shown in Figure
\ref{fig:Figure_correlation_namelen_exception}.
Likewise, some exceptions tend to be more frequent in notebooks with a low \textbf{number of cells}, while others occur in long notebooks about as often as in short ones (cf.\ Figure 
\ref{fig:Figure_correlation_totalcells_exception}).

\begin{figure}[!htb]
\includegraphics[width=1.0\linewidth]{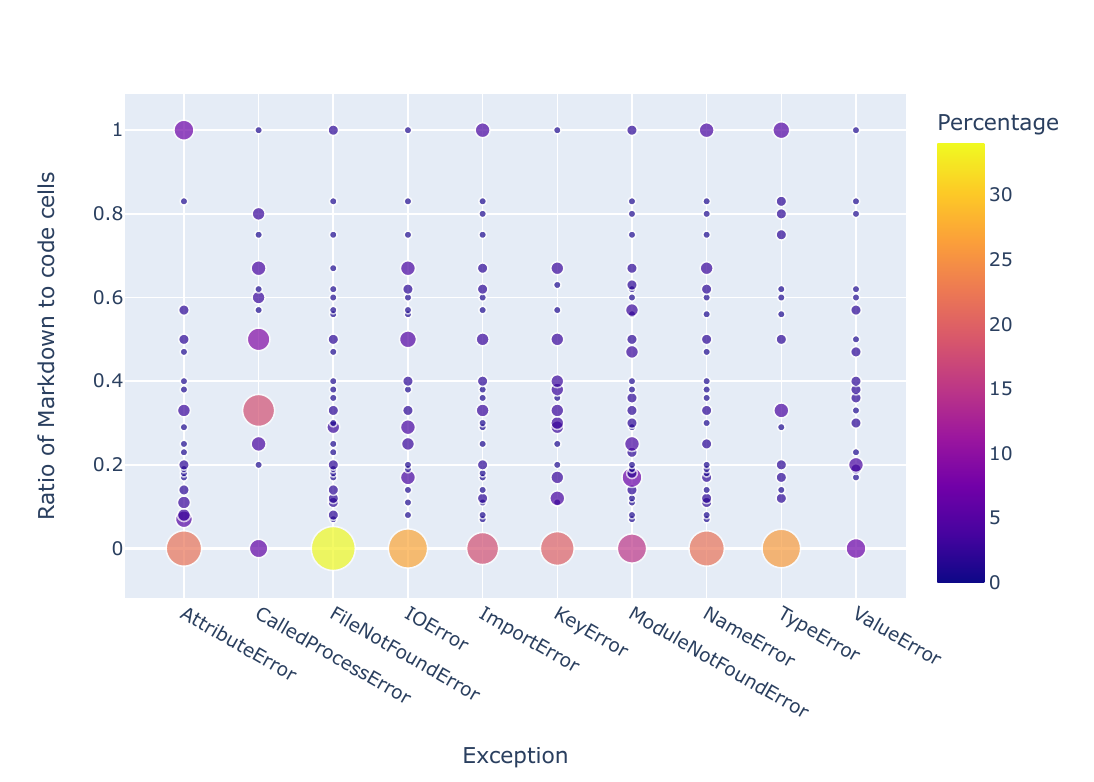}
\caption{
Exceptions by ratio of Markdown to code cells in the corresponding notebooks. 
``Percentage'' represents the percentage of exceptions from notebooks with a given Markdown to code cell value relative to the total number of notebooks associated with that particular exception.
For instance, 34\% of all \textit{FileNoteFoundError} exceptions were due to notebooks with a Markdown to code cell ratio of zero, i.e.\ without any Markdown cells.
}
\label{fig:Figure_correlation_ratio_markdown_code_cells_exception}
\end{figure}

A clearer picture emerges when considering 
the prevalence of exceptions as a function of
the \textbf{Markdown to code cell ratio}, as depicted in Figure
\ref{fig:Figure_correlation_ratio_markdown_code_cells_exception}: low ratios (i.e.\ a relative lack of Markdown cells) correlate with the occurrence of exceptions.


While our study was focused on getting an overview of Jupyter notebook reproducibility in biomedical research, the dataset and methodology
presented here can of course be improved and
used in other contexts. One that we would like to point out here is that of education and training about good computational research practice. Given that the skills required for avoiding~-- or fixing~-- errors vary by error type (Table \ref{tab:error-fixes}),
a dataset like ours that can be queried for notebooks known to cause a specific kind of error can be a useful resource for learners and educators alike when searching for materials that match certain skills. 
The option to filter by additional criteria like Python version, MeSH terms, journal or
article type could be valuable for finding notebooks that match with the interests of the learners, which would increase their motivation to engage with the skills aspects.
We would be happy to collaborate with educators and learners 
to explore how our workflows could be streamlined for such purposes, and we have reached out to some initiatives in this space in order to give this a try.

\subsection{Successful reproductions}
\label{sec:Successful}
1,203 (7.61\%) (\initial{396 (9.50\%)}) of the notebooks in our corpus finished their execution successfully without any errors (cf.\ Figure
\ref{fig:Figure_PRISMA} 
and
Table \ref{tab:comparison}).
However, for 324 (2.05\%) (\initial{151 notebooks (3.62\%)}) of these, our execution generated results that differed from those in the original notebooks, while 879 (5.56\%) (\initial{245 (5.88\%)}) notebooks produced the same results in our execution as documented for the original notebooks.
Of note, the ratio \texttt{different}/(\texttt{different}+\texttt{identical}) changed from 
0.38
in the initial run
to
0.27
in the re-run, indicating that if a notebook ran through, its probability to produce \texttt{identical} results was higher in the re-run than in the initial run, which means more recent notebooks are more likely to yield \texttt{identical}
results.

\begin{figure}[!htb]
\includegraphics[width=1.0\linewidth]{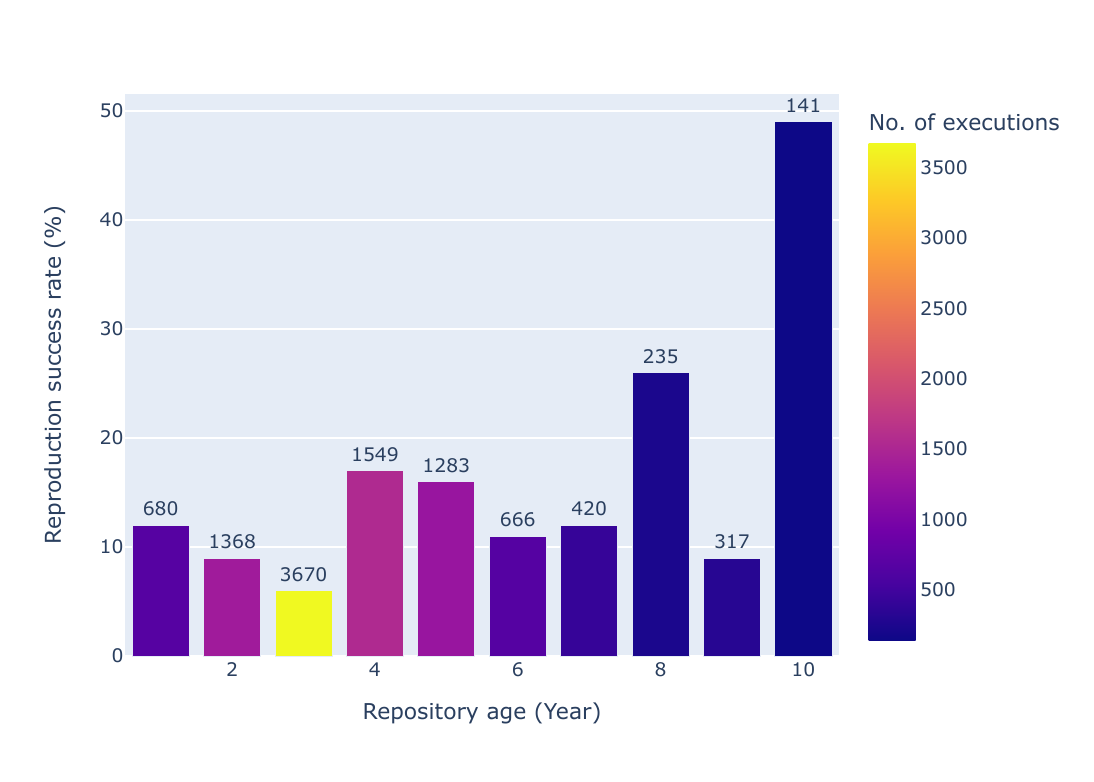}
\caption{
Rate of successful reproduction as a function of the age of the repository (relative to 2023).
On top of the bars is the total number of notebooks per age cohort.
Note that notebooks might be less old than the repository in which they are hosted, but we did not account for that.
}
\label{fig:Figure_decay_rate_replication_success_over_repo_age}
\end{figure}

The relationship between the recency and exceptions is a bit more complex (cf.\ Figure \ref{fig:Figure_decay_rate_replication_success_over_repo_age}), with notebooks from newer repositories not generally performing better than older ones.


%

\begin{figure}[!htb]
    \centering
    \begin{subfigure}[t]{0.46\textwidth}
	\includegraphics[width=1.1\linewidth]{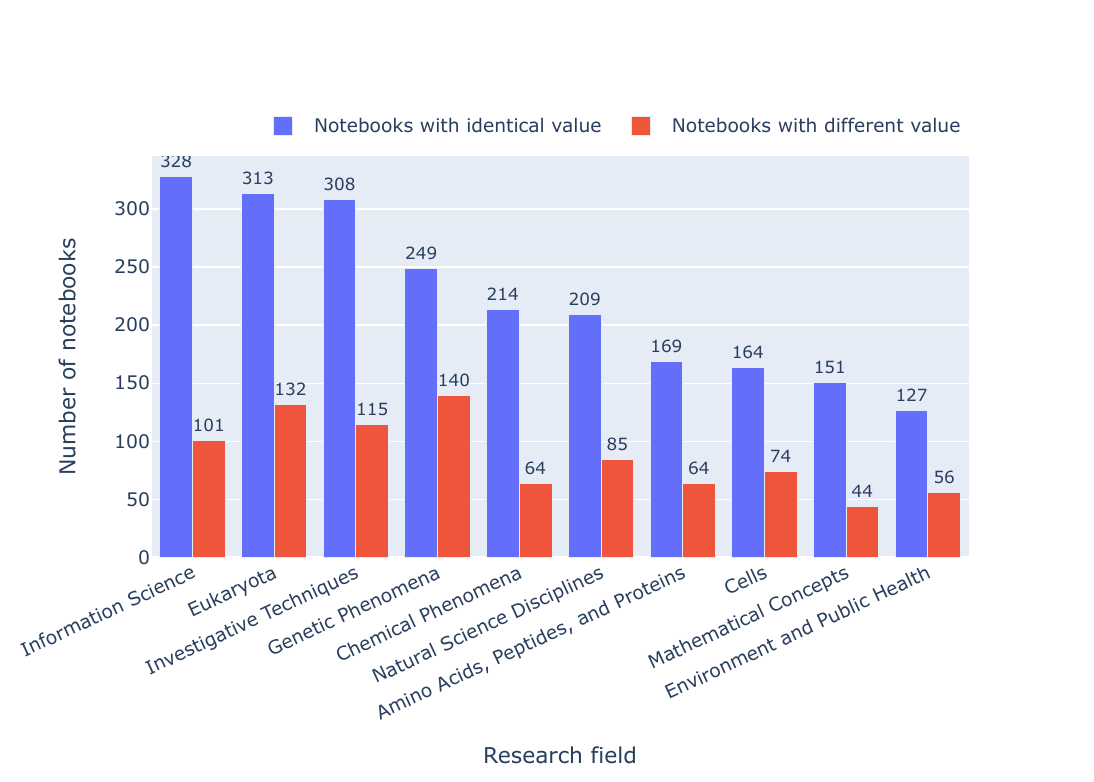}
	\caption{
	Reproducibility of exact results by research field. 
	Considering only notebooks that ran through without triggering an exception, this plot shows the number of notebooks per MeSH term that gave \texttt{identical} (blue) or \texttt{different} (red) results with respect to the originally published notebook.
	In this set of MeSH terms, the ratio \texttt{identical}/(\texttt{identical}+\texttt{different}) was highest for \emph{Mathematical Concepts} and \emph{Amino Acids, Peptides, and Proteins} at ca.\ 0.77 each, and lowest for \emph{Genetic Phenomena} at ca.\ 0.64.
	}
    \label{fig:Figure_notebooks_same_distinct_by_field}
    \end{subfigure}
        \begin{subfigure}[t]{0.48\textwidth}
	\includegraphics[width=1.1\linewidth]{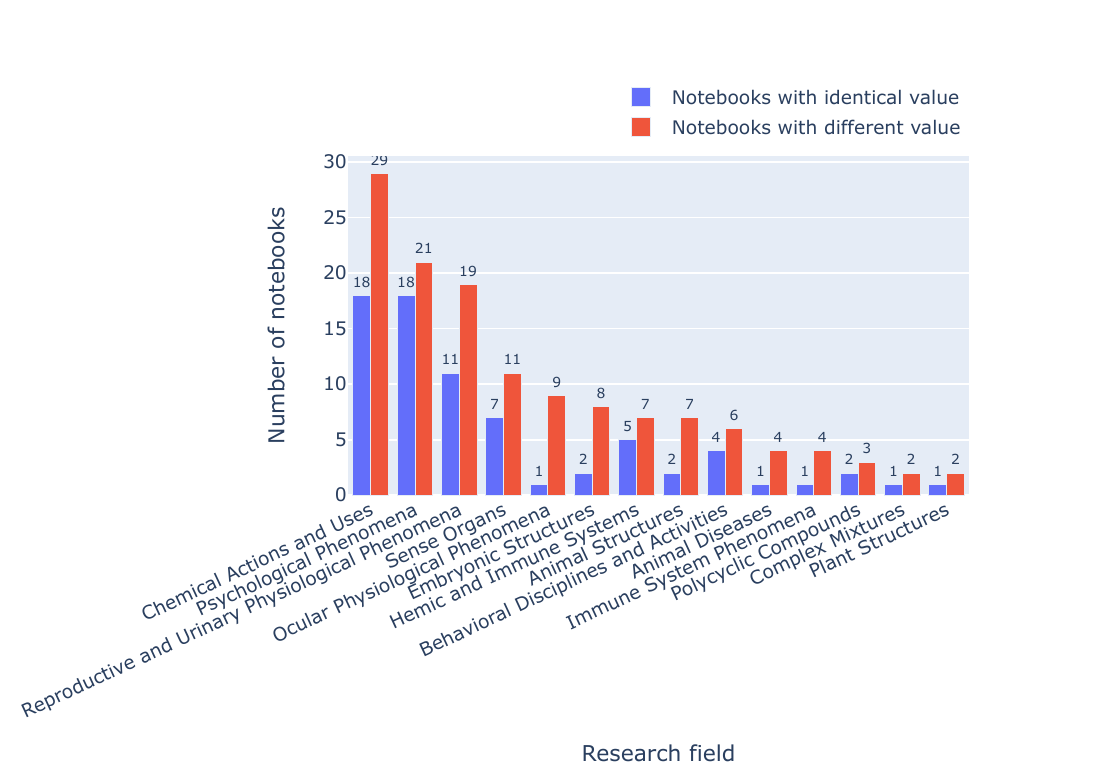}
	\caption{
	Reproducibility of exact results by research field where different results are more frequent than identical ones. 
	In this set of MeSH terms, the ratio \texttt{identical}/(\texttt{identical}+\texttt{different}) was lowest for \emph{Ocular Physiological Phenomena} at 0.1 and highest for \emph{Psychological Phenomena} at ca.\ 0.46.
	}
	\label{fig:Figure_notebooks_distinct_more_than_same_by_field}
    \end{subfigure}
    \caption{
    Reproducibility of notebooks with identical and different results by research field,
    taking upper-level MeSH terms 
    as a proxy.
    }
    \label{fig:Figure_notebook_reproducibility_by_field}

\end{figure}

To get an overview of how different research areas are affected, 
Figure
\ref{fig:Figure_notebook_reproducibility_by_field}
shows the 
number of successful executions of Jupyter notebooks as a function of the MeSH terms for the associated articles, highlighting differences with respect to notebooks that did or did not yield results identical to the ones 
originally reported.
In
Figure
\ref{fig:Figure_notebooks_same_distinct_by_field}, examples are given where \texttt{identical} results were more frequent than \texttt{different} ones,
and in
\ref{fig:Figure_notebooks_distinct_more_than_same_by_field} the inverse.

%

\begin{table}[!htb]
\caption[Comparison of notebooks that were successfully executed without errors]{
Comparison of notebooks that were successfully executed without errors, grouped by whether their results were \texttt{different} from or \texttt{identical} to the results documented for the original notebook. For features listed in \textit{italics}, the mean values per notebook are indicated, otherwise totals across all notebooks per group. 
}
\label{tab:comparison}
\begin{tabular}{| p{0.45\linewidth} | p{0.2\linewidth} | p{0.2\linewidth} |}
\toprule
\textbf{Features} & \textbf{Notebooks with \texttt{different} results}  & \textbf{Notebooks with \texttt{identical} results} \\
\midrule
Number of notebooks     & 324 (\initial {151})    &  879 (\initial{245})   \\
\hline
setup.py     & 0 (\initial{0})    &  344 (\initial{98})   \\
\hline
requirement.txt     & 1 (\initial{0})    &  353 (\initial{107})   \\
\hline
pipfile     & 0 (\initial{0})    &  0 (\initial{0})   \\
\hline
\textit{Total cells}     & 23 (\initial{17.9})    &  19.6 (\initial{17.1})   \\
\hline
\textit{Code cells}    & 15.4 (\initial{12.3})    &  11.3 (\initial{9.8})   \\
\hline
\textit{Markdown cells}    & 7.6 (\initial{5.6})    &  8.3 (\initial{6.7})   \\
\hline
\textit{Ratio of Markdown vs. code cells}    & 0.49 (\initial{0.46})    & 0.73 (\initial{0.68})   \\
\hline
\textit{Empty cells}    & 0.9 (\initial{0.7})    &  0.7 (\initial{0.7})   \\
\hline
\textit{Differences}     & 6.3 (\initial{5.3})    &  0 (\initial{0})   \\
\hline
\textit{Execution time (s)}     & 18.3 (\initial{22.1})    &  57.6 (\initial{16.4})   \\
\hline
\textit{Execution time per code cell (s)}    & 1.88 (\initial{1.80})    &  5.09 (\initial{1.67})   \\

\bottomrule
\end{tabular}
\end{table}

\begin{table}[!htb]
\caption[Comparison of most frequent Python versions declared for notebooks that were successfully executed without errors.]
{
Comparison of most frequent Python versions declared for notebooks that were successfully executed without errors, grouped by whether their results were \texttt{different} from or \texttt{identical} to the results documented for the original notebook. Versions listed in \textit{italics} occur in both top-5 groups, versions listed in \textbf{bold} in only one. The \texttt{count} 
columns give total number of notebooks per version and group, while the \texttt{\%age} columns normalize the absolute values as a percentage of the total number of notebooks per group, i.e. 324 for \texttt{different} and 879 for \texttt{identical}, as per Table \ref{tab:comparison}. In both groups, the top-ranked versions account for slightly over half of the notebooks. }
\label{tab:versioncomparison}
\resizebox{0.48\textwidth}{!}{ 
\begin{tabular}{|r|rrr|lrr|}
\toprule
\multicolumn{1}{|l|}{}     & \multicolumn{3}{c|}{\textbf{\texttt{different}}}                                                                       & \multicolumn{3}{c|}{\textbf{\texttt{identical}}}                                                                      \\ \hline
\multicolumn{1}{|c|}{\textbf{rank}} 
& 
\multicolumn{1}{c|}{\textbf{version}}         
& \multicolumn{1}{c|}{\textbf{count}} & \multicolumn{1}{c|}{\textbf{\%age}} & 
\multicolumn{1}{c|}{\textbf{version}}         
& \multicolumn{1}{c|}{\textbf{count}} & \multicolumn{1}{c|}{\textbf{\%age}} \\
\midrule
1 & \multicolumn{1}{r|}{\textit{3.6}}  & \multicolumn{1}{r|}{184} & 56.8  & \multicolumn{1}{r|}{\textbf{3.7}} & \multicolumn{1}{r|}{457}       & 52\\ 
\hline
2 & \multicolumn{1}{r|}{\textit{2.7}}  & \multicolumn{1}{r|}{90}       & 27.8 & \multicolumn{1}{r|}{\textbf{3.8}} & \multicolumn{1}{r|}{233}       & 26.5\\ 
\hline
3 & \multicolumn{1}{r|}{\textbf{3.4}} & \multicolumn{1}{r|}{25} & 7.7 & \multicolumn{1}{r|}{\textbf{3.9}} & \multicolumn{1}{r|}{125} & 14.2 \\
\hline
4 & \multicolumn{1}{r|}{\textbf{3.5}}  & \multicolumn{1}{r|}{21} & 6.5 & \multicolumn{1}{r|}{\textit{3.6}} & \multicolumn{1}{r|}{50} & 5.7   \\ \hline
5 & \multicolumn{1}{r|}{\textbf{3.1}}  & \multicolumn{1}{r|}{3} & 0.9      & \multicolumn{1}{r|}{\textit{2.7}} & \multicolumn{1}{r|}{10}       & 1.1    \\ 
\bottomrule
\end{tabular}
} 
\end{table}


Table \ref{tab:comparison} zooms in on the successfully executed notebooks 
and compares those that did not yield the same results as the original ones (\texttt{different} group) with those that did (\texttt{identical} group). 
A clear difference between both groups is that many of the notebooks in the \texttt{identical} group had their dependencies specified via either setup.py or requirements.txt or both, in contrast to 
only one
of the notebooks in the \texttt{different} group. 
Since notebooks with no dependency declarations were run using the default conda dependencies, the fact that they successfully finished means that all dependencies were covered. However, as the version of the dependencies used in the original notebook was not documented, it may have differed from the version 
provided in our respective conda environment. 

Besides versioning of dependencies, there could be a number of other reasons as to why an error-free execution might yield different results. For instance, random functions may be invoked, 
dynamic data used,
or code cells in the original might have been executed multiple times or in a different order than in our execution, which ran every code cell just once, from top to bottom.
However, we would not expect 
such circumstances
to correlate so strongly with whether the dependencies had been explicitly declared or not. 

In contrast to 
dependency declarations, other features in Table \ref{tab:comparison} show more gradual differences between the two groups, and some of them fit with intuition. 
For instance, it is understandable that notebooks with more code cells (which is the case for the \texttt{different} group) tend to have a higher probability to yield different results. Likewise, since Markdown cells are indicative of documentation effort, 
notebooks with more Markdown cells (which is the case for the \texttt{identical} group) tend to have a higher probability to yield identical results.
Of particular interest is the ratio of Markdown versus code cells, which is significantly higher for the \texttt{identical} group, which fits with suggestions that it may be a proxy for the notebook quality~\citep{wagemann2022five,venkatesh2023enhancing}, since that ratio is indicative of documentation efforts, and better documentation would be expected to go with better reproducibility.
It would likewise be intuitive to expect that notebooks with more code cells take longer to execute. Indeed, this is what we had observed in the initial run~\citep{samuel2022computational}. Yet in our re-run, the situation was different in that the notebooks 
in the
\texttt{identical} group have fewer code cells but longer total execution times.
This translates into their execution time per code cell being about 2.7-fold of the value of the \texttt{different} group.
We do not have a good explanation for that and invite further research on this.
One could suspect that 
this may reflect the hidden but growing complexity of the code (and data) invoked via the notebooks, including the growing usage of machine learning libraries, though an argument could be made that more complex code raises the probability of different outcomes. 


The average number of differences observed per notebook (or even per code cell) is not easy to interpret on its own, as it includes differences in output cells, cell counter values or in output files, and a difference early in a notebook can lead to further differences later.

Table \ref{tab:versioncomparison} illustrates how different major versions of Python performed in terms of whether successful executions
led to \texttt{identical}
or \texttt{different} results: 
Versions 3.6 and 2.7 were represented 
in both top-5 groups, coming out on top for \texttt{different} 
and at the bottom for 
\texttt{identical}.
The other versions found in the top~5 for \texttt{different}
were Python 3 versions older than 3.6,
while the other versions found in the top~5 for \texttt{identical}
were Python 3 versions newer than 3.6.

\subsection{Notebook styling}
\label{sec:NotebookStyling}

\begin{table}[!htb]
\caption{Common Python code warnings/ style errors
in our notebook corpus. E is for code styling, F for definitions, W for deprecated keys.}
\label{tab:pycodestyling}
\begin{tabular}{| p{0.07\linewidth} | p{0.47\linewidth} | p{0.3\linewidth} |}
\toprule
\textbf{Error code} & \textbf{Description} & \textbf{Count (\%)}\\
\midrule
E231     & missing whitespace after commas, semicolons or colons & 686382 (25.2\%) \initial{102218 (27.3\%)}  \\
E225     & missing whitespace around operator  &  187528 (6.9\%) \initial{25979 (6.9\%)}\\
E265     & block comment should start with ‘\#‘  & 110972 (4.1\%) \initial{10769 (2.9\%)}\\
E402    & module level import not at top of file & 108067 (4.0\%) \initial{10478 (2.8\%)}\\
E262     & inline comment should start with ‘\#‘ & 32704 (1.2\%) \initial{8369 (2.2\%)}\\
E703     & statement ends with a semicolon  & 13944 (0.5\%) \initial{2023 (0.5\%)}\\
E127    & continuation line over-indented for visual indent & 15486 (0.6\%) \initial{1290 (0.3\%)} \\
E701 	& multiple statements on one line & 5147 (0.2\%) \initial{500 (0.1\%)}\\
E741    & do not use variables named ‘l’, ‘O’, or ‘I’ & 2398 (0.1\%) \initial{432 (0.1\%)}\\
E401 	& multiple imports on one line & 1293 (0.0\%) \initial{95 (0.0\%)}\\
E101 	& indentation contains mixed spaces and tabs & 2881 (0.1\%) \initial{32 (0.0\%)}\\
\hline
F405     & \textit{name} may be undefined, or defined from star imports: \textit{module}  & 46825 (1.7\%) \initial{4840 (1.3\%)}\\
F401    & \textit{module} imported but unused & 46988 (1.7\%) \initial{3938 (1.1\%)}\\
F821    & undefined name 'X'  & 20987 (0.8\%) \initial{2071 (0.6\%)}\\
F403    & 'from module import *' used; unable to detect undefined names & 4371 (0.2\%) \initial{263 (0.1\%)}\\
F841    & local variable 'X' is assigned to but never used & 3195 (0.1\%) \initial{225 (0.1\%)}\\
F404    & future import(s) name after other statements & 309 (0.0\%) \initial{44 (0.0\%)}\\
F402    & import 'X' from line Y shadowed by loop variable & 79 (0.0\%) \initial{10 (0.0\%)}\\ 
F633    & use of >> is invalid with print function & 6 (0.0\%) \initial{6 (0.0\%)}\\
F823    & local variable 'X' defined in enclosing scope on line Y referenced before assignment & 12 (0.0\%) \initial{4 (0.0\%)}\\
\hline
W601 	& .has\_key() is deprecated, use ‘in’ & 27 (0.0\%) \initial{7 (0.0\%)}\\
W606    & 'async' and 'await' are reserved keywords starting with Python 3.7 & 3 (0.0\%) \initial{3 (0.0\%)}\\
\bottomrule
\end{tabular}
\end{table}

In addition to the common exceptions,
we also checked the notebooks for code styling errors, as shown in
Table \ref{tab:pycodestyling}, which presents the error code for the Python code warnings and style errors found in our study.
E231 is the most common coding style error, followed by E225 and E265, respectively.
There are also some common content errors other than styling errors like F403 and F405~-- these are related to variable and module definition errors.
The W601 and W606 warnings relate to the use of deprecated and reserved keys.

While the results are similar overall for both the initial run and the re-run, a few minor differences can be observed: the relative prevalence of E231 (whitespace) and E262 (comments) has decreased slightly, while that of E265 (comments), E402 (module import), E127 (indentation), F405 (name/ module) and F401 (module) has increased. 


\begin{table}[!bh]
\caption{
Common types of exceptions encountered in Python-based Jupyter notebooks in our corpus (ordered as per Figure \ref{fig:Figure_top_exception_by_reason}), 
along with notes on their nature and a brief outline of how they can be addressed (other than by verifying the spelling of the respective commands). }
\label{tab:error-fixes}
\begin{tabular}{| L{0.3\linewidth} | L{0.28\linewidth} | L{0.28\linewidth} |}
\toprule
\textbf{Error type} & \textbf{Underlying problem} & \textbf{Some potential fixes} \\
\midrule
ModuleNotFoundError & module can not be located                                                                                               & check that module is present in repo or installed in the environment                                                           \\
\midrule
FileNotFoundError   & file cannot be located at designated path                                                                                       & check path and that file exists at path                                                                                        \\
\midrule
ImportError         & attribute, function, class, or variable cannot be imported from a module as specified                                                 & check documentation of what is to be imported, including the module's dependencies                                             \\
\midrule
NameError           & variable or function is used but not defined                                                                            & check documentation about where it ought to be defined (e.g. another cell or module); execute that code before using that name \\
\midrule
IOError             & trying to read from or write to a destination that does not exist or for which user does not have pertinent permissions & check existence, path and permissions of that destination                                                                      \\
\midrule
AttributeError      & trying to access an attribute or method that does not exist as specified                                                & check documentation and ensure the access is handled as required                                                               \\
\midrule
ValueError          & a function is called with an argument of the correct type but with a wrong value                                        & check that argument meets the requirements of the function                                                                     \\
\midrule
TypeError           & a function is called with an argument of the correct type but with a wrong value                                        & check that argument number and argument types meet the requirements                                                            \\
\midrule
KeyError            & trying to access a dictionary key that does not exist                                                                   & check that the key exists in the target dictionary; consider setting default values for cases when key does not exist          \\
\midrule
CalledProcessError  & a subprocess was called but returned a non-zero exit status                                                             & check that the subprocess is being called as required and that the called code actually works as intended \\   \bottomrule
\end{tabular}
\end{table}



\subsection{Environmental footprint} 
\label{sec:Environmental-footprint-results}
For the initial run of the pipeline, we obtained an estimate of 47.38 kWh. Using the default values for Germany, this means an approximate carbon footprint of 16.05 kg CO2e, which is equivalent to 17.51 tree months.
For the re-run, 
the pipeline consumed 373.78 kWh, resulting in a carbon footprint of approximately 126.58 kg CO2e, equivalent to 11.51 tree years when using default values for Germany.
Our hardware had 18 cores per CPU, and the footprint calculation accounted for that, though
our code did not have provisions for running on multiple cores. We do not have detailed information on whether more than one core was actually used but the \textit{multiprocessing} module, for instance~-- one of the libraries commonly used for multi-core processing~-- was present in 
338 notebooks in our corpus, so we can assume it was used when called before the first exception or in notebooks that ran through.






\section{Discussion}
\label{sec:Discussion}

In this study, we have analyzed 
the \textit{Method reproducibility}~--~in the sense of \citet{goodman2016what}~--~
of Jupyter notebooks written in Python and publicly hosted on GitHub that are mentioned in publications whose full text was available via PubMed Central by 
the day when our reproducibility pipeline was started,  i.e.\ on 27 March 2023 (\initial{24 February 2021}). We will now contextualize some aspects of the study and 
then discuss its limitations as well as implications, again primarily for \textit{Method reproducibility} 
of Jupyter notebooks associated with biomedical publications.

\subsection{Contextualization}
\label{subsec:Contextualization}

\subsubsection{Exploring interactions between Jupyter and research via Wikidata}
\label{sec:Wikidata}






In research contexts like those investigated here,
Jupyter notebooks are often used alongside other resources, which may be software, data, instruments, physical materials, mathematical models and so forth~-- all of which affect scientific reproducibility.
Our pipeline captured only some facets of that
but the relationships between Jupyter notebooks and 
various
aspects of the research ecosystem~--
as
highlighted, for instance, in 
Figures
\ref{fig:Figure_top_researchfields_with_articles},
\ref{fig:Figure_top_journals_with_articles},
\ref{fig:Figure_top_notebook_language} and
\ref{fig:Figure_f_notebook_module_full_import}~--
partly overlap with what can be explored via Wikidata,
a cross-disciplinary and multilingual database through which a global community curates FAIR and open data to serve as general reference information \citep{waagmeester2020wikidata,rutz2022LOTUS}. This includes
data
about key elements of the research ecosystem, from researchers to research fields and research organizations, from methods to datasets, software and publications.

While coverage and annotation of the scholarly literature in Wikidata are far from complete, some initiatives focused on research software in particular have begun to explore Wikidata as a space to curate information 
related to
software in research contexts \citep{rasberry2022scholia,levitskaya2022analysis,istrate2022large}.
Once integrated into Wikidata, such software-related information can be explored in various ways that combine the software and the non-software parts of the Wikidata knowledge graph. A popular option to do that is through the visualization tool Scholia \citep{nielsen2017scholia,rasberry2022scholia}, which provides profiles for different types of entities or relationships. 
\begin{figure}
    \centering
    \includegraphics[width=\linewidth]{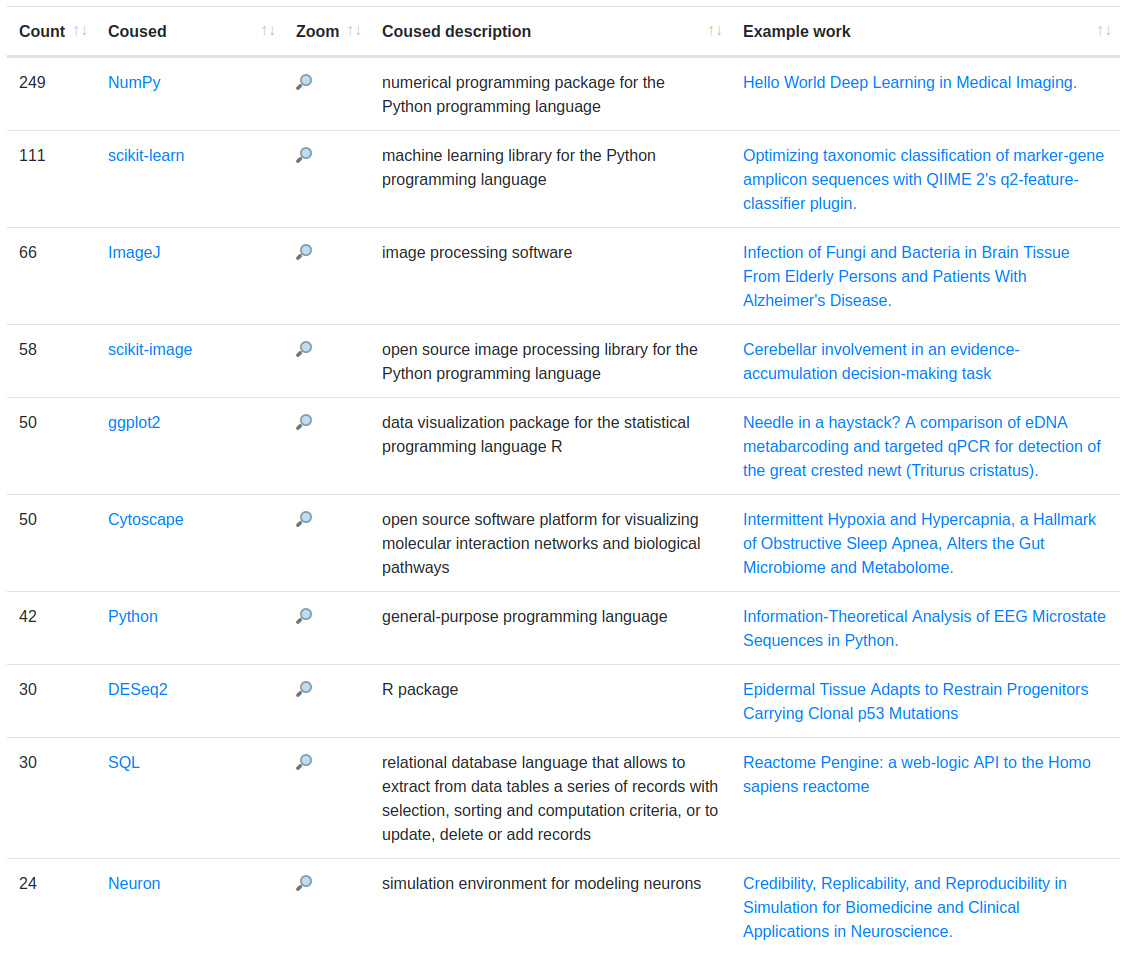}
    \caption{Scholia panel from the \textit{use} profile for Jupyter notebook, 
    displaying the results of a Wikidata query for research resources commonly used together with Jupyter notebooks. The magnifying glasses link to \textit{uses} profiles that display information about co-use of the respective research resource alongside Jupyter notebooks.}
    \label{fig:Wikidata-Scholia}
\end{figure}

For 
Jupyter notebooks
(known to Wikidata as \href{https://www.wikidata.org/wiki/Q70357595}{Q70357595}), 
the most relevant profile types 
in Scholia
are those for a research \textit{topic}\footnote{\href{https://scholia.toolforge.org/topic/Q70357595}{https://scholia.toolforge.org/{\textbf{topic}/Q70357595}}} 
(portraying, e.g., studies, people and venues related to research about Jupyter notebooks),
a \textit{software}\footnote{\href{https://scholia.toolforge.org/software/Q70357595}{https://scholia.toolforge.org/\textbf{software}/Q70357595}} (portraying, e.g., software dependencies)
or a \textit{use}\footnote{ \href{https://scholia.toolforge.org/use/Q70357595}{https://scholia.toolforge.org/\textbf{use}/Q70357595}} (portraying, e.g., studies and people using Jupyter notebooks).
The \textit{use} profile, for instance, features a panel with examples of research resources used alongside the resource being profiled. This panel is show in part in 
Figure \ref{fig:Wikidata-Scholia}
%
for 
Jupyter notebooks.
Besides
Python packages like \textit{numpy} and \textit{scikit-learn} (similar to Figure \ref{fig:Figure_f_notebook_module_full_import}), it also shows non-Python software like 
\textit{DESeq2} (an R package often run via Jupyter environments) or 
\textit{ImageJ} (written in Java and run outside Jupyter) or non-software items like \textit{Bayes' theorem} or \textit{10x Genomics Chromium}. 
%
These co-usages can be explored further via dedicated \textit{uses} profiles linked from that panel's entries\footnote{\href{https://scholia.toolforge.org/uses/Q70357595,Q197520}{https://scholia.toolforge.org/\textbf{uses}/Q70357595,Q197520}}.
Some profile types can be combined, e.g. the \textit{organization}  profile for the
European Molecular Biology Laboratory (EMBL)\footnote{\href{https://scholia.toolforge.org/organization/Q1341845}{https://scholia.toolforge.org/\textbf{organization}/Q1341845} }
has a \textit{use} panel whose Jupyter entry links to the profile of EMBL-associated scholarship using Jupyter notebooks\footnote{\href{https://scholia.toolforge.org/organization/Q1341845/use/Q70357595}{https://scholia.toolforge.org/\textbf{organization}/Q1341845/\textbf{use}/Q70357595}}.


Although incomplete in its coverage of the research literature in general and biomedical publications in particular, 
Wikidata does cover publications and software across many research fields. Since anyone can edit it, its coverage of any particular aspect~-- say, 
reproducibility\footnote{\href{https://scholia.toolforge.org/topic/Q1425625}{https://scholia.toolforge.org/topic/Q1425625}}
or the demographics of GitHub contributors~\citep{levitskaya2022analysis}~--
can be improved as needed. To assist with that, Scholia provides curation pages for most of its profile types\footnote{\href{https://scholia.toolforge.org/use/Q70357595/curation}{https://scholia.toolforge.org/\textbf{use}/Q70357595/\textbf{curation}}}.

%






\subsubsection{Uptake dynamics of Jupyter notebooks parallel those of ORCID} 
\label{sec:Uptake}

\begin{figure}[!htb]
\includegraphics[width=1.0\linewidth]{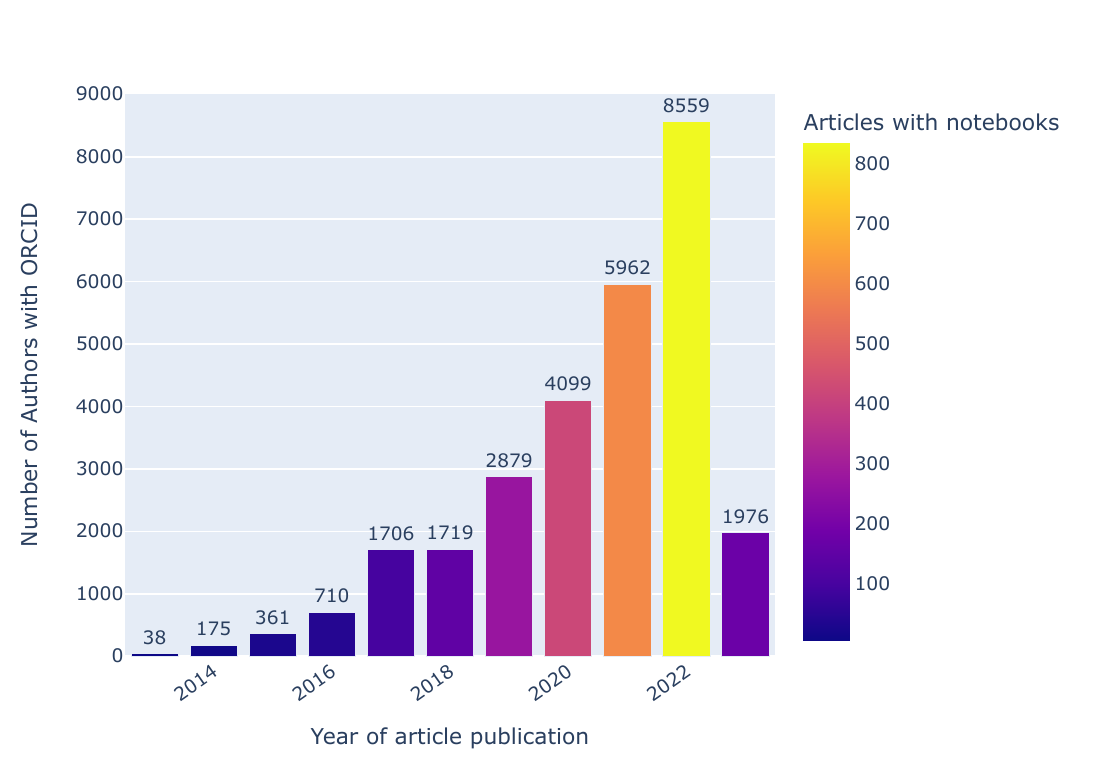}
\caption[ORCID usage in our collection.]{
ORCID usage in our collection. Bars indicate the total number of ORCIDs found each year for authors of articles in our collection. Colors indicate the number of articles that year with Jupyter notebooks. Note that data for 2023 are incomplete.
}
\label{fig:Figure_timeline_articles_authors_with_orcid}
\end{figure}

As part of our exploration of the broader research landscape around 
Jupyter notebooks, we analyzed the uptake of ORCID identifiers\footnote{\url{https://orcid.org/}} over time in the collected journal articles with notebooks (Figure \ref{fig:Figure_timeline_articles_authors_with_orcid}).
ORCID provides a persistent digital identifier to uniquely identify authors and contributors of scholarly articles~\citep{haak2018using}. 
While IPython notebooks go back to 2001, the Jupyter notebooks with kernels for multiple languages became available in 2014~\citep{kluyver2016jupyter}, 
whereas ORCID was launched in 2012~\citep{Haak2012ORCID}. Hence, both are relatively recent innovations in the scholarly communications ecosystem, and their respective uptake processes occur in parallel. 

In 2017, there were 98 Jupyter notebooks associated with articles in our corpus, versus 833 in 2022
(cf.\ Figure \ref{fig:Figure_timeline_articles_with_without_notebooks}), which means a growth by about an order of magnitude over the course of five years.
Over a similar time span, the number of ORCIDs found each year for authors of articles in our collection grew 
by about an order of magnitude too, from 710 in 2016 to 8,559 in 2022 
(cf.\ Figure \ref{fig:Figure_timeline_articles_authors_with_orcid}).

\subsection{Limitations}
\label{subsec:Limitations}

The present study does not address \textit{Inferential reproducibility} and only briefly touches upon  \textit{Results reproducibility}. Furthermore, we made no attempt to re-run
computational notebooks that met any of the following exclusion criteria during the reference period: (a) they did not use Jupyter (or its precursor, IPython), (b) notebooks written in languages other than Python or not using the Python kernel, 
(c) they were not publicly available on GitHub, (d) they were not mentioned in publications available from PubMed Central, (e) they were not on the base branch of their GitHub repository (which is the only branch we looked at).

Our reproducibility workflow is based on that from \cite{pimentel2019a},
with some changes to include GitHub repositories from publications and using the \textit{nbdime} library 
\citep{nbdime}
from Jupyter instead of string matching for finding differences in the notebook outputs.
The approach is using conda \citep{Conda}
environments. 
We considered only the first error for any given notebook~-- there may be additional ones, and they may or may not depend on the first one.
We did not use any Docker images \citep{Docker2013}
for the execution environment, even in cases when they may have been available.
Currently, the pipeline exclusively parses AST to gather details about modules, statements, expressions, etc. However, this data isn't employed for installing dependencies. This offers a potential path for future exploration, involving an extended approach that considers not just the provided requirement files, but also docker files, conda environment specifications (e.g., env.yml), and import statements to construct an execution environment for notebook execution.
We looked at Markdown cells as a proxy for documentation effort but did not look into the use of comments in code cells.
We did not make any adjustments for code that is supposed to be run on multiple cores 
or on GPUs or TPUs, and we did not record how many of the 18 cores of our system were actually used.

For a good number of the reported problems (especially the missing software or data dependencies, as per Figure \ref{fig:Figure_top_exception_by_reason}),
it is often straightforward to fix them manually for individual notebooks (e.g.\ as per Table \ref{tab:error-fixes}), yet undertaking manual fixes systematically was not practical at the scale of the thousands of notebooks rerun here, and designing a pipeline for automated fixes (e.g.\ as per \citep{wang2021restoring}) was out of scope. 
That said, 
\citep{woodbridge2017jupyter} 
reports on a manual fixing attempt (which also provided the foundation for a prototypical automated notebook validation tool that makes use of GitLab Actions\footnote{\url{https://gitlab.com/mwoodbri/jupyter-ci}}), while
\citep{schroder2019reproducible} examined 22 notebooks from five PMC-indexed publications in detail, including with some attempted manual fixes. 

If the original code had specified dependencies without referring to a specific version, our rerun would use the most recent 
conda-installable
version of that library.
Another important aspect here is that of community engagement, e.g.\ authors of notebooks could be contacted systematically and asked for their input on how they already deal with reproducibility issues, and where they see room for improvements, including in light of our findings.
We assessed the age of repositories rather than specifically measuring the age of individual notebooks (see Figure \ref{fig:Figure_decay_rate_replication_success_over_repo_age}).

Finally, in estimating the environmental footprint of this study, we only included the footprint due to running the full pipeline once~-- we did not include the efforts involved in preparing and testing the pipeline, analyzing the data or writing the manuscript.



\subsection{Implications}

There are several implications of this study, and we welcome collaborations around any of them.

First, on a general level, the low degree of reproducibility that we documented here for Jupyter notebooks associated with biomedical publications 
goes conform with similarly low levels of reproducibility that were found in earlier domain-generic studies, both for Python \citep{rule2018exploration,schroder2019reproducible,pimentel2021understanding}
and R
\citep{trisovic2022large}.
This is a problem that needs further attention, particularly from users and providers of computational and related resources.

Second, considering that the notebooks we explored here were associated with peer-reviewed publications, it is clear that the review processes currently in place at journals within our corpus does not generally pay much attention to the reproducibility of the notebooks, though our data indicates that quality gradients can be observed, e.g., 
by research field
(cf.\ Figures
\ref{fig:Figure_notebooks_same_distinct_by_field} and 
\ref{fig:Figure_notebooks_distinct_more_than_same_by_field}), 
journal (cf.\ Figure \ref{fig:Figure_exceptions_by_journal_notebooks_by_percentage}) or article type (cf.\ Figure
\ref{fig:Figure_exceptions_by_subject_article_normalized}).
%
%
%
%
%
This clearly needs to improve, and while 
Figure 
\ref{fig:Figure_notebooks_same_distinct_by_field} and
the peaking exception rate in Figure 
\ref{fig:Figure_timeline_exceptions_by_year_notebook_scatter_percentage}
are positive signs,
we need systemic approaches to that rather than just adding this to the list of things the reviewers are expected to attend to. As our study demonstrates, a basic level of reproducibility assessment can well be achieved in a fully automated fashion, so it would probably be beneficial in terms of research quality to include such automated basic checks~-- for notebooks and other software~-- into standard review procedures. Ideally, this would be done in a way that works across publishers as well as for a variety of technology stacks and programming languages. Additional provisions~-- e.g.\ for sampling subsets if the amount of data or the time required for 
reproducing
the original computations exceed certain limits~-- might be useful too.

Third, 
while there is a large variety in the types of errors affecting reproducibility, some of the most common errors concentrate around dependencies (cf.\ Figure \ref{fig:Figure_f_notebook_module_full_import},
\ref{fig:Figure_f_notebook_module_load_ext_toplevel}, \ref{fig:Figure_dependencies} and \ref{fig:Figure_top_exception_by_reason}), so efforts aimed at systemic improvements of dependency handling~-- e.g.\ as per \cite{zhu2021restoring}~--
have great potential to increase reproducibility. 
Here, programming language-specific efforts regarding code dependencies can be combined with efforts targeted at improving the automated handling of data dependencies, which would be beneficial irrespective of the specific programming language.
Researchers attempting to publish research with associated notebooks should not have to do this all by themselves~-- research infrastructures as well as publishers and funders can all help 
establish 
best practice 
and engaging communities around that.
Despite its small scale of only
12 articles published so far\footnote{\url{https://elifesciences.org/collections/d72819a9/executable-research-articles}}, 
the \textit{Executable Research Articles} 
initiative\footnote{\href{https://elifesciences.org/labs/ad58f08d/introducing-elife-s-first-computationally-reproducible-article}{https://elifesciences.org/labs/ad58f08d/introducing-elife-s-first-computationally-reproducible-article}} at 
\textit{eLife} is interesting in this regard.
However, 
the lack of new additions for over a year
render it
unclear at this point how robust, scalable and maintainable the underlying technology stack is.

Fourth, zooming in on Python specifically, wider adoption of existing workflows for code dependency management (such as \textit{requirements.txt}, \textit{conda} environment files, or \textit{Poetry}\footnote{\url{https://python-poetry.org/}}) would help, and so would 
standardized checks~-- of dependencies, versions, executability and output validity~-- during the publishing process
(cf.\ Table \ref{tab:comparison}). 
 
Fifth, the few notebooks that actually did reproduce (cf.\  \nameref{sec:Successful})
are not equally distributed (cf.\ Figures \ref{fig:Figure_notebook_reproducibility_by_field},
\ref{fig:Figure_exceptions_by_journal_notebooks_by_percentage} and
\ref{fig:Figure_exceptions_by_subject_article_normalized}).
This means that reproducibility could probably be strengthened by enhancing or highlighting  the features that correlate with it. For instance, Jupyter notebooks with higher documentation effort generally scored better than others (cf.\ Table \ref{tab:comparison}), underlining once more the importance of documentation~\citep{napflin2019genomics,leipzig2021role}.
In more specific terms, it also seems worthwhile to have a closer look at 
the workflows for creating, documenting, reviewing and publishing notebooks associated with
journals 
like \textit{iScience} (cf.\ Figure \ref{fig:Figure_exceptions_by_journal_notebooks_by_percentage})
or with article types like 
\emph{Tools and Resources}
(cf.\ Figure \ref{fig:Figure_exceptions_by_subject_article_normalized}).
Furthermore, there is merit in the idea of making Jupyter notebooks or similar environments for combining computational and narrative elements
a publication type of their own. This is already the case in some places, as examplified by \cite{constantine2016python} or \cite{garg2022pygetpapers} in the \textit{Journal of Open Source Software}.

Sixth, the ongoing diversification of the Jupyter ecosystem~-- e.g.\ in terms of programming languages, deployment frameworks or cloud infrastructure~--
is increasingly reflected, albeit with delay, in the biomedical literature. In parallel, while GitHub remains hugely popular, alternatives like GitLab, Gitee or Codeberg are growing too. 
Future assessments of Jupyter reproducibility will thus need to take this increasing complexity into account, and ideally present some systematic approach to it.

Seventh, the delays that come with current publishing practices also mean that Jupyter notebooks associated with freshly published papers are using software versions near or even beyond their respective support window (which is 42 months in much of the Python ecosystem\footnote{Cf.\ \href{https://numpy.org/neps/nep-0029-deprecation_policy.html}{https://numpy.org/neps/nep-0029-deprecation\_policy.html}}).
For instance, the oldest Python version still officially supported in 2023 was 3.7, which was sunset on 27 June 2023\footnote{See \href{https://endoflife.date/python}{https://endoflife.date/python} for release schedule}),
yet as shown in Figure
\ref{fig:Figure_timeline_python_minor_version_by_repo_update}, 
about four thousand Python notebooks from repositories whose last commit was in 2023 still featured earlier Python versions, mainly 
3.6 (sunset in 2021) but also 
2.7 (2020), 3.5 (2020), 3.4 (2019) and some for which the version could not be determined. This contributes to reproducibility issues.
A similar issue exists with the versions of the libraries called from any given notebook, though the effects might differ as a function of whether they have been invoked with or without the version being specified. If the version had been specified, its official end of life might go back even further. If the version was not specified, the newest available version would be invoked, which may not be compatible with the way the library had been used in the original notebook. 
Similar issues can arise with the versioning of APIs, datasets, ontologies or other standards used in the notebook, all of which can contribute to reduced reproducibility.
To some extent, these version delay issues can be shortened by preprints: since they are (essentially by definition, but not always in practice) published before the final version of the associated manuscript, and hence their delays should be shorter, with lower reductions in reproducibility, though we did not investigate that in detail. 

Eight, the variety (cf.\ 
Table \ref{tab:error-fixes})
and scale (cf.\ Figure \ref{fig:Figure_exceptions_by_field}) of issues encountered in the notebooks analyzed here provide ample opportunities for use in educational contexts~-- including instructed, self-guided or group learning~-- since fixing real-life errors (cf.\ Table \ref{tab:error-fixes}) or warnings (cf.\ Table \ref{tab:pycodestyling}) can be more motivating than working primarily with textbook examples.
To do this effectively would require some mapping of the strengths and weaknesses of the notebooks to learning objectives
or curriculum requirements,
which may range from understanding programming paradigms, software engineering principles or data integration workflows to developing an appreciation for documentation and other aspects of good scientific practice\citep{sayres2018bioinformatics}. 
Given the continuously expanding breadth of publications that use Jupyter notebooks, it is also steadily becoming easier to find publications where they have been used in research meeting specific criteria. These could be a particular topic~-- 
e.g.\
natural products research \citep{mayr2020finding}
or invasion biology \citep{bors2019population}~-- or workflows involving a particular experimental
methodology
like single-cell RNA sequencing \citep{vargo2020rank} or other software tools like ImageJ \citep{bryson2020composite}.
It is already possible to query our dataset for articles with a specific MeSH term and associated notebooks with a specific type of exception or with 
replication status \texttt{different}.
We are exploring how our materials and workflows and the insights derived from them can be integrated 
with educational initiatives like The Carpentries
\citep{Wilson2014Software,pugachev2019what}.

%

Ninth, our analysis identified 879 notebooks for which we have documented reproducibility in terms of obtaining the same results as in the original study.
What mechanisms should be used~-- 
and at what level (e.g.\ article, repository or notebook, or aggregations of any of these)~--
to communicate this kind of reproducibility to the scientific community? Badges could be an option, and they have had some effect in related circumstances~\citep{hardwicke2021analytic,cruwell2023whats},
but it would not be clear what social processes should be used for awarding, displaying or otherwise handling them.
Dedicated reproducibility platforms like ReScience~\citep{rougier2017sustainable}\footnote{\url{https://rescience.github.io/}} work fine for reproducibility studies 
at the level 
of individual notebooks or small numbers, but it is not clear how they would handle the scales discussed here.
Nanopublications are another option, and they too have been experimented with in related settings~\citep{bucur2023nanopublication}.
While it is perhaps relatively uncontroversial to mark successful reproducibility of a given resource, 
what 
should be done
about the 324 notebooks that gave different results, or about all the others that raised exceptions, had installation problems or a missing GitHub repository?
We are interested in exploring these issues in order to 
increase the impact of reproducibility studies and the reusability of
their results .
%

Tenth, 
our corpus and methodology could be useful in terms of bringing guidance for good computational practice closer to the actual workflows. 
We are thus working towards distilling the insights from this study into recommendations and infrastructure that assist with making Jupyter notebooks more reproducible and facilitate validation of some basic levels of reproducibility.

\section{Conclusions}
\label{sec:Conclusions}

On the basis of re-running 
15,817 (\initial{4,169})
Jupyter notebooks associated with 
3,467 (\initial{1,419})
publications whose full text is available via Pubmed Central, we conclude that
such notebooks are becoming more and more popular for sharing code associated with biomedical publications, that the range of programming languages or journals they cover is continuously expanding
and that their reproducibility is low but improving,
consistent with earlier studies on Jupyter notebooks shared in other contexts.

The main issues are related to dependencies~-- both code and data~-- which means that reproducibility could likely be improved considerably if the code~-- and dependencies in particular~-- were better documented. Further improvements could be expected if some basic and automated reproducibility checks of the kind performed here
were to be systematically included in the peer review process or if computational notebooks~-- Jupyter or otherwise~-- were combined  
 with additional approaches that address reproducibility from other angles, e.g. registered reports.


\section{Data availability}
\label{sec:Dataavailability}
All the data generated during the initial study can be accessed at \url{https://doi.org/10.5281/zenodo.6802158} \cite{samuel2022Dataset}, while the data from the re-run is available at \url{https://doi.org/10.5281/zenodo.8226725} \cite{samuel2023Dataset}.
The code used is available at \url{https://github.com/fusion-jena/computational-reproducibility-pmc}.
The code contains notebooks used for analysis of the results.

\subsection{Ethical Approval (optional)}
No facet of the research reported here triggered a requirement for ethical review. While our data contains personally identifiable information, it was taken directly from PMC.
We did, however, consider the ethical implications of automated reproducibility studies of the kind presented here, which led us to (a) highlight systemic aspects, (b) not zoom in on individual notebooks or their authors
and (c) include environmental footprint information.

\subsection{Consent for publication}


Not applicable.

\subsection{Competing Interests}
The authors declare there are no competing interests.

\subsection{Funding}


Work by S.S. was supported 
by the Carl Zeiss Foundation for the project ``A Virtual Werkstatt for Digitization in the Sciences (K3)'' \citep{samuel2020virtual} within the scope of the program line ``Breakthroughs: Exploring Intelligent Systems for Digitization - explore the basics, use applications''. Work by D.M. was supported by the Alfred P. Sloan Foundation under grant number G-2021-17106 \citep{rasberry2022scholia} and by the MaRDI project~\cite{MaRDIProject} under DFG grant number 460135501.
The computational experiments were performed on resources of Friedrich Schiller University Jena supported in part by DFG grants INST 275/334-1 FUGG and INST 275/363-1 FUGG.

\subsection{Author's Contributions}


S.S conceived and designed the experiments, performed the experiments, analyzed the data, prepared figures and/or tables, authored or reviewed drafts of the paper, and approved the final draft.

D.M. conceived and designed the experiments, performed the experiments, analyzed the data, prepared figures and/or tables, authored or reviewed drafts of the paper, and approved the final draft.

\section{Acknowledgements}

We would like to thank the providers of infrastructure, data and code that we used in this study. These include the PubMed Central repository hosted by the National Center for Biotechnology Information in the United States and the Ara Cluster at the University of Jena as well as the Python, Jupyter and Conda communities and their respective dependencies.
We acknowledge the Open Research Doathon on the occasion of Open Data Day 2017, where the first attempts at systematic reproduction of PMC-indexed Jupyter notebooks were made \cite{woodbridge2017jupyter}.
Special thanks go to JupyterCon, which made the two of us aware of each other's work and provided the nucleus for our collaboration.

\section{Authors' information (optional)}


S.S is a computer scientist working in e-Science, Semantic Web, and Machine Learning. Her research focuses on enhancing the reproducibility of scientific studies by leveraging provenance, linked data, and knowledge graphs, with an aim to fostering transparency and facilitating the sharing and reuse of research data.

D.M. is a biophysicist working on integrating open research and education workflows with the web, for all stages of the research cycle. His research activities span across the spatial and temporal scales of life and concentrate on the data aspects of research, particularly in the life sciences and adjacent fields. 

\bibliography{main}

\end{document}